\documentclass[useAMS,usenatbib]{mn2e}

\usepackage{graphicx}

\title[X-ray calibration of the Y-$M_{tot}$ scaling law with ACCEPT]{X-ray calibration of SZ scaling relations with the ACCEPT catalogue of galaxy clusters observed by \textit{Chandra}}
\author[]{B. Comis$^{1}$\thanks{E-mail: barbara.comis@roma1.infn.it}, M. De Petris$^{1}$, A. Conte$^{1}$, L. Lamagna$^{1}$ and S. De Gregori$^{1}$\\
$^{1}$Sapienza University of Rome, Department of Physics, P.le A. Moro, 2
Rome, Italy}
\begin{document}

\date{Accepted for publication in MNRAS}

\pagerange{\pageref{firstpage}--\pageref{lastpage}} \pubyear{2011}
\maketitle

\label{firstpage}
\begin{abstract}
We explore the scaling relation between the flux of the Sunyaev-Zel'dovich (SZ) effect and the total mass of galaxy clusters using already reduced Chandra X-ray data present in the ACCEPT (Archive of Chandra Cluster Entropy Profile Tables) catalogue. The analysis is conducted over a sample of 226 objects, examining the relatively small scale corresponding to a cluster overdensity equal to 2500 times the critical density of the background universe, at which the total masses have been calculated exploiting the hydrostatic equilibrium hypothesis. Core entropy ($K_{0}$) is strongly correlated with the central cooling time, and is therefore used to identify cooling-core (CC) objects in our sample. Our results confirm the self-similarity of the scaling relation between the integrated Comptonization parameter ($Y$) and the cluster mass, for both CC and NCC (non-cooling-core) clusters. The consistency of our calibration with recent ones has been checked, with further support for $Y$ as a good mass proxy. We also investigate the robustness of the constant gas fraction assumption, for fixed overdensity, and of the $Y_{X}$ proxy \citep{K07} considering CC and NCC clusters, again sorted on $K_{0}$ from our sample.  We extend our study to implement a $K_{0}$-proxy, obtained by combining SZ and X-ray observables, which is proposed to provide a CC indicator for higher redshift objects. Finally, we suggest that an SZ-only CC indicator could benefit from the employment of deprojected Comptonization radial profiles.
\end{abstract}


\section{Introduction}
The Sunyaev-Zel'dovich effect is produced by the interaction of cosmic microwave background (CMB) photons with the high-energy electrons of the ICM (intra-cluster medium) \citep{SZ, Sun80, Rephaeli, Birk}. The effect produces a distortion of the CMB spectrum and could therefore provide a redshift independent probe of the pressure of the electron population that produced it. 
Within this perspective the study of the relations between the flux of the SZ effect and other observables becomes of the utmost interest, providing alternative tools for investigating cluster evolution up to large $z$, and so back in time. 

When integrating the SZ flux over the whole cluster extent we actually measure the total thermal energy of the ICM, and so the underlying gravitational potential. The integrated Comptonization parameter $Y$, proportional to the SZ flux, could therefore be a strong, low scatter mass proxy, as confirmed both by numerical simulation works \citep[e.g.][]{daSilva04, Motl, K07, Aghanim} and by joint SZ/X-ray analyses \citep[e.g.][]{Reid, Nagai06, B2008, Huang, Sayers}. 

In our work we present an X-ray only calibration of the $Y-M_{tot}$ scaling relation, using 226 clusters observed by the \textit{Chandra X-ray Observatory} and collected into the ACCEPT catalogue \citep{Cavagnolo09}. The analysis has been conducted considering the cluster mean overdensity $\Delta$ equal to 2500 times the critical density of the background universe, at the cluster redshift. We have thus chosen to probe scales that should be more sensitive to non-gravitational processes and for which X-ray data could be more suited than simulations and SZ observations.   
The main interest of such calibrations has always focused on obtaining X-ray predictions of SZ-flux scaling at larger radii \citep[][]{Vikhlinin09, Arnaud}, better matched to the resolution and beam size of present-day SZ instruments. However, the increasing number of current SZ studies that use X-ray calibrated $Y-M_{tot}$ relations to iteratively estimate cluster masses at $\Delta = 500, 200$ \citep{Andersson, PlanckPratt} pose the need to verify whether the self-similar trend is preserved even for smaller radii. 
No other X-ray calibration can be found in literature for the $\Delta=2500$ overdensity. At these small scales, joint SZ/X-ray analyses have already been performed by \cite{Morandi07}, considering 24 non-uniformly observed clusters, and by \cite{B2008}, exploiting the isothermal $\beta$-model for the ICM of 38 objects (33 of which are also ACCEPT clusters), divided into two sub-sets in $z$. Both studies  have considered SZ projected signals, thus probing the cylindric volume through the object, and compared them to spherically integrated masses. Exploiting X-ray data, we here calibrate the scaling law for quantities integrated over spherical volumes. The self-similarity of the low scatter $Y$-$M_{tot}$ relation has been tested on the basis of a strong indicator of ICM physics, the core entropy excess $K_{0}$, made available by \cite{Cavagnolo09} for this catalogue. The larger data-set allowed us to explore the redshift dependence of the scaling parameters over a higher number of $z$-sorted subsamples. A redshift dependent bias could in fact be very dangerous for cosmological applications of the considered scaling law and of extreme interest to constrain the different processes that determine the evolution of galaxy clusters \citep{Nath}.

With a view to the incoming results of SZ blind surveys, the quest for an observable able to distinguish between CC and NCC clusters even at large redshift becomes of great interest. For this reason several studies \citep{Santos, Hudson, Pipino} have suggested possible CC-proxies based on existing X-ray data-sets. 
We propose a core entropy proxy constructed from the central Comptonization parameter $y_{0}$ and the central electron number density $n_{e,0}$, in order to show a $K_{0}$ indicator independent of any spectroscopic information and temperature profile. An SZ only CC identifier is also considered, by investigating the correlation between $K_{0}$ and the ratio of SZ signals integrated within different spherical volumes. In fact, it would be very useful to identify a possible SZ-only cooling core indicator, in order to exploit redshift independence of the effect.

The paper is organized as follows: In Section \ref{scaling_rel} we introduce the SZ scaling relations, as expected in the self-similar scenario. The data-set and the models used to analytically represent the ICM are discussed in Section \ref{Data_set}. Section \ref{scatter} is dedicated to scaling law analysis, while further core entropy proxies are introduced in Section \ref{CC_det}. Finally, the main results and conclusions are summarized in Section \ref{conclusion}.

The whole work has been developed assuming a flat-$\Lambda CDM$ universe, with the following choice of cosmological parameters: $\Omega_{M}$=0.3, $\Omega_{\Lambda}=0.7$ and $h=0.7$ ($H_{0}=100$ $h$ $kms^{-1}Mpc$)

\section{Thermal SZE: scaling relations and self-similarity}\label{scaling_rel}
In ionized regions, such as the ICM, energy transfer from electrons to CMB photons could take place because of the thermal SZ effect. The amplitude of the effect is proportional to the Comptonization parameter $y$ that, assuming spherical symmetry for the cluster, is given by
\begin{equation}
	y(\theta)=\int^{\infty}_{0}\frac{k_{B}T_{e}}{m_{e}c^{2}}\sigma_{T} n_{e}dl,
\end{equation}
at each projected angular distance from the cluster center $\theta=r/D_{A}$ ($D_{A}$ being the angular diameter distance of the cluster).
$y(\theta)$ is determined by the line-of-sight integral of the electron pressure, product of the electron gas temperature $T_{e}$ and the gas number density $n_{e}$ ($m_{e}$ is the electron rest mass, $c$ the light speed, $k_{B}$ the Boltzmann constant, $\sigma_{T}$ the electron Thomson scattering cross section). 
The integrated Comptonization parameter $Y$ is obtained by integrating $y(\theta)$ over the solid angle $\Omega$ subtended by a projected area on the sky ($A$),
\begin{equation}
	Y=\int_{\Omega}yd\Omega=\frac{1}{D_{A}^{2}}\frac{k_{B}\sigma_{T}}{m_{e}c^{2}}\int^{\infty}_{0}dl\int_{A}n_{e}T_{e}dA.
\label{Y_int}
\end{equation}

Very simple scaling relations can be obtained considering the self-similar scenario \citep{Kaiser}. In our study the cluster evolution is assumed to be self-similar with respect to  critical density of the universe  at each cluster redshift $\rho_{c}(z)=3H_{0}^{2}E(z)^{2}/8\pi G$, where $E(z)^{2}=\left(H(z)/H_{0}\right)^{2}=\Omega_{M}(1 + z)^{3} + \Omega_{\Lambda}$ accounts for the background universe evolution and $G$ is the gravitational constant. We have then fixed the overdensity $\Delta$, which defines the scale radius $r_{\Delta}=\left(3 M_{tot,\Delta}/4\pi\Delta \rho_{c}\right)^{1/3}$. 

The virial theorem relates the total mass to the electron gas temperature $T_{e}$ as 
\begin{equation}
	k_{B}T_{e}=\mu m_{p}\frac{GM_{tot}(r<R)}{2 R}
\end{equation}
($\mu$ is the mean molecular weight of the gas and $m_{p}$ the proton mass). To calculate $Y_{S}$ and $Y_{C}$ respectively, probing the spherical and the cylindric projected volumes, equation (\ref{Y_int}) can be rewritten directly in terms of the electron pressure profile $P_{e}(r)$:
\begin{equation}
	Y_{S,\Delta}=4\pi \frac{\sigma_{T}}{m_{e}c^{2}}\int^{r_{\Delta}}_{0}P_{e}(r)r^{2}dr,
	\label{Ys}
\end{equation}
\begin{equation}	
	Y_{C,\Delta}=\frac{2\pi}{D_{A}^{2}}\frac{\sigma_{T}}{m_{e}c^{2}}\int^{r_{\Delta}}_{0}\left(2 \int^{\infty}_{0}P_{e}(r)dr\right) r dr,
	\label{Yc}
\end{equation}
with $Y_{S,\Delta}=Y_{C,\Delta}D_{A}^{2}/C$ (where $C$ accounts for the different integration domains).
The total gas mass within a radius $r_{\Delta}$ is calculated by integrating the gas density profile $\rho_{gas}(r)=\mu_{e}m_{p}n_{e}(r)$:
\begin{equation}
	M_{gas,\Delta}=4\pi\mu_{e}m_{p}\int^{r_{\Delta}}_{0}n_{e}(r)r^{2}dr
	\label{Mgas}
\end{equation}
(with $\mu_{e}$ the electron mean molecular weight). It can be related to the total mass through the gas fraction $f_{gas}$, $f_{gas,\Delta}=M_{gas,\Delta}/M_{tot,\Delta}$.
Then the $Y-M_{tot}$ scaling relation can be obtained by combining the previous equations
\begin{equation}
	Y_{S,\Delta}=\frac{\sigma_{T}}{m_{e}c^{2}}\frac{\mu}{\mu_{e}}\left(\frac{\sqrt{\Delta} G H_{0}}{4}\right)^{2/3}E(z)^{2/3} f_{gas,\Delta}M_{tot,\Delta}^{5/3}.
	\label{eq_scaling}
\end{equation}
In this paper we only consider the overdensity $\Delta=2500$, and we hereafter omit the subscript (e.g. $M_{tot}=M_{tot,2500}$).

\section{Sample, ICM modelization and analysis}\label{Data_set}
\subsection{Data-set}\label{data_c09}
Our analysis has considered the cluster sample of the ACCEPT project, produced by \cite{Cavagnolo09} (\textit{C09} hereafter). This catalogue\footnote{http://www.pa.msu.edu/astro/MC2/accept/} provides the scientific community with homogeneously observed and reduced public X-ray data taken with \textit{Chandra}, covering a wide range of cluster redshift (z $\sim 0.05-0.89$), bolometric luminosity ($L_{X}\sim10^{42-46}erg/s$) and temperature ($T_{X}\sim 1-20keV$). 

This \textit{Chandra} archival project collects clusters observations performed for many different programs. The high number of objects included, accounting for a variety of cluster morphologies, should ensure that the catalogue is not unbalanced, oversampling a particular class of clusters, even if neither flux nor volume limited. Therefore the data-set can be considered nearly unbiased, in particular for the applications presented in this work (see also Section \ref{entropy}). Our results should also remain unaffected by the overestimation of the CC fraction, typical of X-ray observations.

The quality of \textit{Chandra} data enabled \textit{C09} to produce the observed radial profiles of gas density and temperature from which entropy has been also derived. We selected 226 objects from the complete catalogue. Nine clusters were excluded because their deprojection required a single $\beta$-model fit to the X-ray surface brightness, instead of the concentric annuli approach used to obtain the observed radial profiles of all the other objects \citep[see][APPENDIX A]{Cavagnolo09}. Six further clusters were omitted since their density and pressure profiles were too shallow to constrain our parametrized ICM model (described in the following sections).
For a deeper discussion about data reduction we refer to \cite{Cavagnolo08} and references contained therein. 

\subsection{Modeling the ICM}
The choice of model adopted for our analysis follows that of \cite{M09}, in order to provide an accurate representation of both CC and NCC objects \citep{M09, Conte}. The analytic model for the electron pressure profile is that proposed by \cite{Nagai06} (N07), while the electron gas density is described considering a simplified version of the model suggested by \cite{Vikhlinin} (SVM).

The N07 parametrized pressure profile is a generalization of the NFW model:
\begin{equation}
	P_{e}(r)=\frac{P_{e,i}}{\left(\frac{r}{r_{p}}\right)^{c}\left[1+\left(\frac{r}{r_{p}}\right)^{a}\right]^{\left(b-c\right)/a}}.
\end{equation}
The parameters $a$, $b$, $c$ are respectively the slopes for intermediate radii ($r\sim r_{p}$, a scale radius), the outer region ($r\gg r_{p}$) and the core region ($r\ll r_{p}$), $P_{e,i}$ is a scalar normalization of the pressure profile. As discussed in \cite{Arnaud}, if ICM observations are limited to $r<r_{500}$, the outer slope $b$ is substantially unconstrained by the data and could therefore give rise to important uncertainties in the profile extended to larger radii. Our sample provides pressure profiles up to $\sim r_{2500}$ then we fixed $a=0.9$ and $b=5.0$, allowing only $c$ to be constrained by the data\footnote{The following tern [a, b, c] =[0.9, 5.0, 0.4] \citep[see][for a N07 errata]{M09} provides a good description of the pressure profiles for the observed \textit{Chandra} clusters (at $r\leq r_{500}$)  as well as for the simulated sample (0.5$<r/r_{500}<$2.0).}.

The electron density radial profile has been modeled exploiting the SVM,
\begin{equation}
	n_{e}(r)=n_{e0}\left[1+\left(\frac{r}{r_{c}}\right)^{2}\right]^{-3\beta/2}\left[1+\left(\frac{r}{r_{s}}\right)^{\gamma}\right]^{-0.5\epsilon/\gamma}.
\end{equation}
The further term in addition to the traditional $\beta$-model is able to describe the steepening at $r\sim r_{s}$, the scale radius of the change in slope, while $r_{c}$ still represents the core radius. The parameter $\gamma$ accounts for the width of the transition region between the two profiles and is kept $\gamma=3$, a value that closely matches all the clusters considered by \cite{Vikhlinin}, who constrained $\epsilon$ to be $<5$, to avoid unphysical profiles, as discussed by the authors.

\begin{figure}
\centering
\includegraphics[width=0.8\columnwidth]{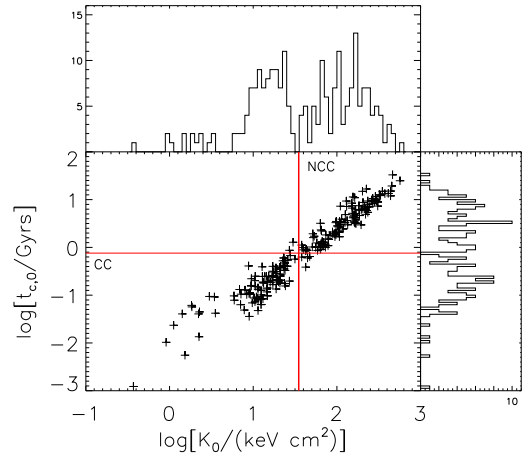}
  \caption{\itshape Strong correlation between $K_{0}$ and $t_{c,0}$. The red solid lines (at $K_{0}=35 keV cm^{2}$) show the lack of clusters mentioned in the text. The corresponding threshold in central cooling time is at $t_{c,0}\sim0.7 Gyr$.}
  \label{fig_tc0_K0}
\end{figure}

\subsection{ICM entropy}\label{entropy}
The ICM entropy $K(r)=k_{B}T_{X}(r)/n_{e}(r)^{2/3}$ represents a more significant probe of the ICM than density and temperature as considered separately \citep[further details can be found in the review by][]{Voit}, and therefore turns out to be a very useful quantity for analyzing the breaking of self-similarity.

Observations \citep[\textit{C09, }][]{Pratt} show that cluster entropy profile flattens at small radii and could be represented as
 \begin{equation}
 K(r)=K_{0}+K_{100}\left(\frac{r}{100 kpc}\right)^{\alpha},
 \end{equation}
$K_{0}$ being the so called core entropy, $K_{100}$ a normalization at 100 kpc and $\alpha$ the power law index.
The power law index is characterized by a small dispersion around the self-similar expected value \citep[$K(r)\propto r^{1.1}$,][]{Tozzi}. By contrast, the entropy floor $K_{0}$ spans a wider range of values and is characterized by a bimodal distribution \citep[\textit{C09, }][]{Pratt, Hudson}. 

\textit{C09} find a weakly populated region in the $K_{0}\simeq30-50$ $keV cm^{2}$ range, as shown by Figure \ref{fig_tc0_K0}, in which we plot the histogram of $K_{0}$ best-fitting values (as obtained by \textit{C09}) for the 226 objects of the ACCEPT collection used in our study. According to \textit{C09}, the same bimodal behavior is also kept for the flux limited subsample of the HIFLUGCS clusters, 94\% of which is present within the ACCEPT catalogue. Therefore, $K_{0}$ distribution cannot be simply attributed to a selection bias. Entropy profile flattening observed towards the cluster center could not even be explained by the effect of PSF systematics. The authors state that the instrument resolution could only have produced a selection effect for clusters with $K_{0}<10$ $keV$ $cm^{2}$, that are not found at $z>0.1$. This should not affect our results, as it we be will discussed later in Section \ref{Y_rel}.

$K_{0}$ is intimately related to the cooling timescale in the cluster core $t_{c,0}$, as expected and also verified by \textit{C09}, exploiting the approximation of \cite{Donahue05}.
Figure \ref{fig_tc0_K0} highlights the strong correlation expected between $K_{0}$ and $t_{c,0}$ for the 226 objects considered in the present work. Core entropy can therefore be exploited as an identifier of CC-objects \citep{Hudson}. 
In the following we will study SZE scaling relations by dividing our data-set into clusters with $K_{0} \le 35$ $keV cm^{2}$ (CCs) and $K_{0} > 35$ $keV cm^{2}$ (NCCs). 

\begin{figure*}
\centering
\begin{minipage}[!b]{\textwidth}
\includegraphics[width=0.9\textwidth]{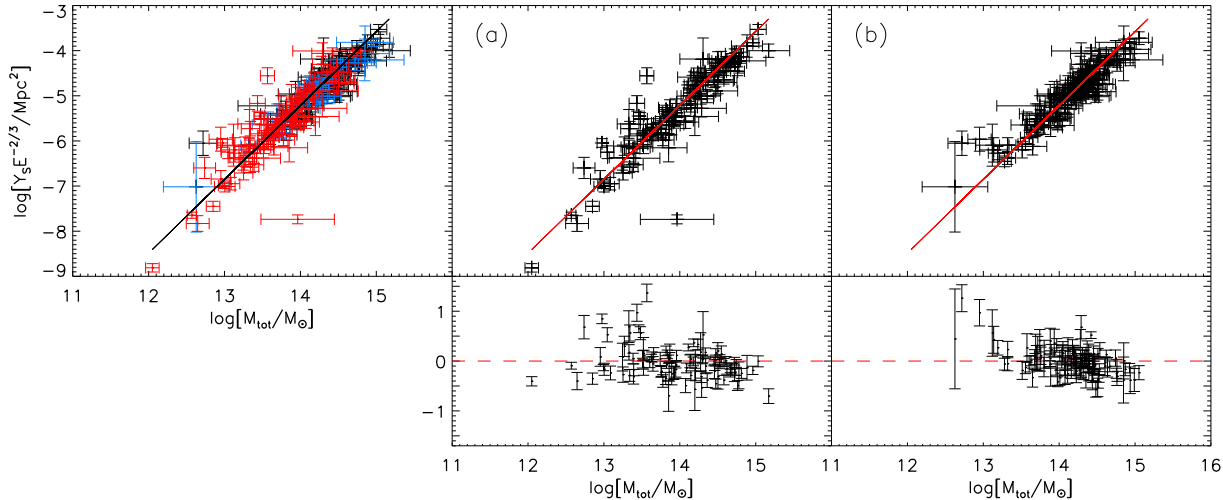}\\
  \caption{\itshape Logarithmic plot of $Y_{S}$ as a function of the cluster total mass $M_{tot}$; the solid black line represents the best-fitting scaling relation (whose parameters are listed in Table \ref{M_Ys_param}). Different redshift subsamples are reported in different colors (black: $z \ge 0.2$, blue: $0.1 < z < 0.2$, red: $z \le 0.1$). Panel $(a)$ shows the $K_{0} \le 35 keV cm^{2}$ subsample, the $K_{0} > 35 keV cm^{2}$ is represented instead on the right panel $(b)$. For both panels, the solid red line represents the best-fitting trend obtained considering the whole sample. At the bottom of plots $(a)$ and $(b)$, deviations from the best-fitting relation are reported, in order to provide a graphic representation of the intrinsic scatter relevance.}\label{fig_YsphMtot}
\end{minipage}

\end{figure*}
\begin{table*}
\centering
\caption{Best-fitting parameters of the $Y-M_{tot}$ scaling relation. The quantities $\sigma_{raw}$ and $\sigma_{int}$ are always calculated with respect to the best-fitting relation obtained considering the whole sample of objects (\textit{all}).}
\label{M_Ys_param} \centering
\renewcommand{\footnoterule}{}  
  \begin{tabular}{ | c c | c c c c |}
    \hline
    $K_{0}$ & $$ & $A$ & $B$ & $\sigma_{raw}$ & $\sigma_{int}$\\
    \hline
    \hline
    $all$ & \tiny{$Y_{S} [Mpc^{2}] - M_{tot} [M_{\odot}]$} & $1.637\pm0.062$ & $-28.13\pm0.88$ & $0.34$ & $0.23$\\
    $all$ ($C_{r}\le1.1$) & \tiny{$Y_{C}D_{A}^{2} [Mpc^{2}] - M_{tot} [M_{\odot}]$} & $1.60\pm0.10$ & $-27.4\pm1.4$ & $ $ & $ $\\
    \hline
    $K_{0}\le35 keV cm^{2}$ & \tiny{$Y_{S} [Mpc^{2}] - M_{tot} [M_{\odot}]$} & $1.650\pm 0.082$ & $-28.3\pm1.1$ & $0.37$ & $0.30$\\
    $K_{0} > 35 keV cm^{2}$ & \tiny{$Y_{S} [Mpc^{2}] - M_{tot} [M_{\odot}]$} & $1.579\pm 0.091$ & $-27.4\pm1.3$ & $0.30$ & $0.13$\\
    \hline
  \end{tabular}
\end{table*}

\begin{table*}
\centering
\caption{Comparison of our $Y-M_{tot}$ calibration with previous works, as listed in the first column. $A^{*}$ and $B^{*}$ are the best-fitting parameters reported in the papers listed, $\Delta$ specifies the overdensity considered for the best-fitting values in the corresponding line. Instead $A$ and $B$ are the results of the calibration performed in the present paper, opportunely converted to the units of measurement and overdensities chosen in the studies we refer to in each line.}
\label{M_Ys_param_others} \centering
\renewcommand{\footnoterule}{}  
  \begin{tabular}{ | l l | c c c c c|}
    \hline
    $$ & $$ & $\Delta$ & $A^{*}$ & $B^{*}$ & $A$ & $B$\\
    \hline
    \hline
    \textit{\cite{B2008}} & \tiny{$Y_{C}D_{A}^{2} [Mpc^{2}] - M_{tot} [M_{\odot}]$} & $2500$ & $1.66\pm0.20$ & $-28.23\pm3.00$ & $1.60\pm0.10$ & $-27.4\pm1.4$\\
    \textit{\cite{Andersson}} & \tiny{$Y_{S} [M_{\odot}keV] - M_{tot} [3\cdot 10^{14}M_{\odot}]$} & $500$ & $1.67\pm0.29$ & $14.06\pm0.10$ & $1.637\pm0.062$ & $14.18\pm0.45$\\
    \textit{\cite{PlanckPratt}} & \tiny{$Y_{S} [Mpc^{2}] - M_{tot} [6\cdot 10^{14}M_{\odot}]$} & $500$ & $1.72\pm 0.08$ & $-4.183\pm0.013$ & $1.637\pm0.062$ & $-4.16\pm0.12$\\
    \hline
    \textit{\cite{Arnaud}} & \tiny{$Y_{S} [M_{\odot}keV] - M_{tot} [3\cdot 10^{14}M_{\odot}]$} & $500$ & $1.790\pm0.015$ & $14.11\pm0.03$ & $1.637\pm0.062$ & $14.18\pm0.45$\\
    \hline
  \end{tabular}
\end{table*}

\subsection{Derived cluster global quantities}\label{mcmc}
In order to obtain the best-fitting parameters for ICM analytical profiles, we used a Markov Chain Monte Carlo (MCMC) technique, using the Metropolis algorithm while including the convergence test proposed by \cite{GRtest}. We improved the efficiency of the method by including the information about the correlations into the proposal \citep{Hanson}. For both the $n_{e}(r)$ and the $P_{e}(r)$ profiles we worked with the log likelihoods, $ln\left(\mathcal{L}\right)\propto -\chi^{2}$.

The choice of $\Delta=2500$ guarantees, for the whole sample, that the integration radius is of the order of the maximum radius observed and allows us to probe scales that are expected to be more sensitive to the contribution of radiative processes.

Under the spherical symmetry assumption, the total mass at a given radius $r$ has been calculated from the hydrostatic equilibrium equation
\begin{equation}
	M_{tot}(r)=-\frac{r^{2}}{G\mu m_{p}n_{e}(r)}\frac{dP_{e}(r)}{dr},
	\label{HE}
\end{equation}
given the gas pressure $P(r)=\mu_{e}P_{e}(r)/\mu$. An iterative procedure is developed to find $r=r_{2500}$, at which the gas mass (eq. \ref{Mgas}) and  the integrated Comptonization parameters (eq. \ref{Ys} and eq. \ref{Yc}) have also been calculated.

\section{Scaling relations analysis}\label{scatter}
Simple power laws can be studied through a linear fit in the log space, $log(Y)=A log(X)+B$, where $A$ and $B$ are the parameters to be estimated. In the present work the log-log fit has been performed by using an MCMC approach, working with log likelihoods, by defining
\begin{equation}
	\chi^{2}=\sum\frac{(log(Y_{i}) - B - Alog(X_{i}))^{2}}{\sigma^{2}_{log(Y_{i})} + (B\sigma log(X_{i}))^{2}}.
\end{equation}

In order to	distinguish the intrinsic dispersion (i.e. due to peculiar ICM physics) around the best-fitting trend we follow the same approach adopted by \cite{ohara} and \cite{PlanckPratt}. The so called \textit{raw scatter} is given by the error-weighted residuals
\begin{equation}
	\sigma_{raw}^{2}=\frac{1}{N-2} \Sigma_{i=1}^{N}w_{i}\left(Y_{i}-A X_{i}-B\right)^{2},
\end{equation}
with $w_{i}=N\sigma_{i}^{-2}/\Sigma_{i=1}^{N}\sigma_{i}^{-2}$ and $\sigma_{i}^{2}=\sigma_{Y_{i}}^{2}+A^{2} \sigma_{X_{i}}^{2}$, expressed as $log(observable)$. For each point we add in quadrature a constant value to $\sigma_{i}$ in order to find the $\sigma$ values so that $\chi^{2}/\nu$ ($\nu$ being the degrees of freedom) becomes equal to 1 for the considered relation.

\subsection{Y - M$_{tot}$ scaling relation}\label{Y_rel}
We have addressed the $Y-M_{tot}$ scaling law (equation \ref{eq_scaling}) taking into consideration the 226 objects of our data-set (Figure \ref{fig_YsphMtot}). 

The analysis was conducted exploiting the widely adopted assumption of a constant gas fraction for the whole sample, and $f_{gas}$ was then included into the normalization of the scaling law. The two $K_{0}$ sorted subsamples were also considered and reported in panels $(a)$ and $(b)$ of Figure \ref{fig_YsphMtot}. In Table \ref{M_Ys_param} we list the best-fitting values obtained for the $Y-M_{tot}$ scaling relation, all estimated by performing a \textit{delete-d Jackknife} resampling, in order to exclude a sample dependence of the results. The values obtained for slope $A$ in the different cases are all consistent with the self-similar expected trend and with each other. 
The CC population seems to show a steeper relation, but there is no significant evolution with $K_{0}$, in agreement with existing literature, from \cite{Morandi07} to the most recent \cite{PlanckPratt}. The intrinsic scatter, with respect to the best-fitting scaling, obtained for the whole sample, is larger for the $K_{0}\le35 keV cm^{2}$ clusters, even if it does not completely dominate the raw scatter (Table \ref{M_Ys_param}), as can also be deduced by looking at bottom panels of Figure \ref{fig_YsphMtot} (a) and (b).  The integrated Comptonization parameter is then a robust mass proxy, not very sensitive to ICM physics even at the relatively small radius $r_{2500}$.

\begin{figure}
\centering
\begin{minipage}[!b]{\columnwidth}
\includegraphics[width=\columnwidth]{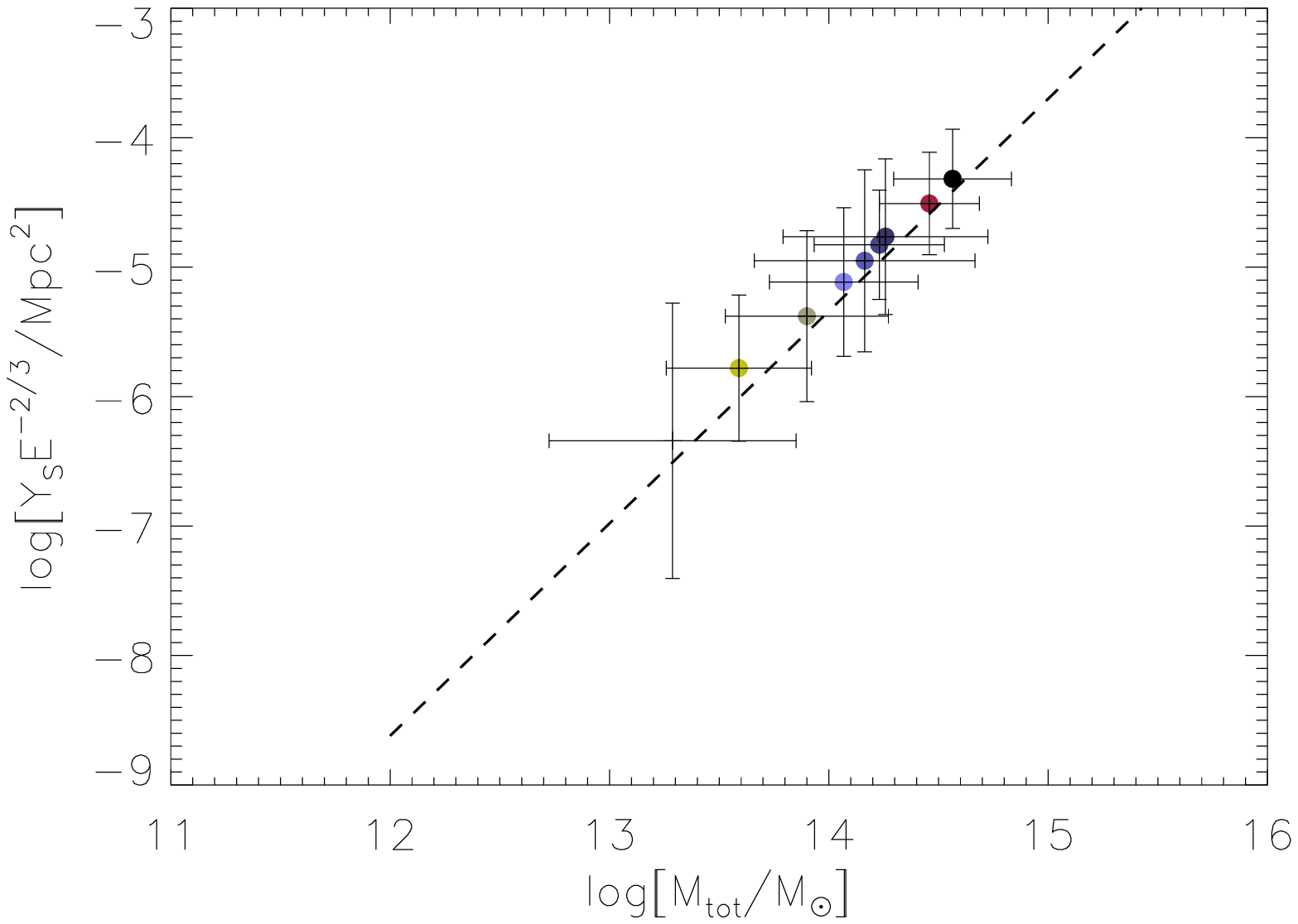}\\
\includegraphics[width=\columnwidth]{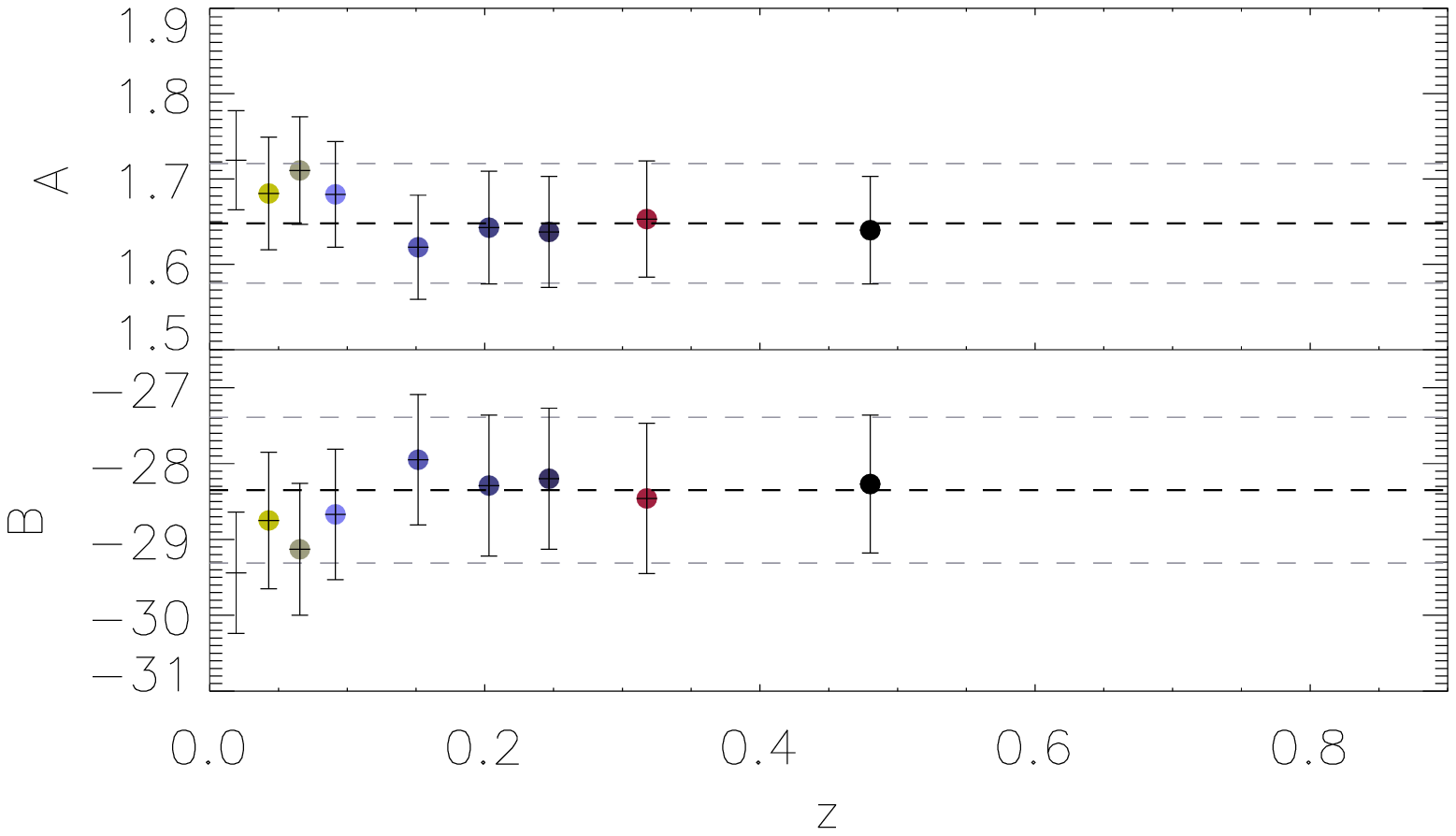}\\
  \caption{\itshape Top: $Y_{S}$ vs $M_{tot}$ for the average values of the nine $z$-sorted subsamples (from white to dark for increasing redshift). The plot shows no evolution with $z$, and the MCMC approach with a self similar prior returns $A=1.648\pm0.070$ and $B=-28.35\pm0.96$ (dashed line). Bottom: Considering the same prior, we show the best-fitting parameters obtained for each sub-set as a function of its mean redshift. The black dashed line is the best-fitting scaling drawn in the upper plot (in grey the $\pm 1 \sigma$ uncertainty).}\label{fig_YsphMtot_z_bin}
  \end{minipage} 
\end{figure}

\begin{figure}
\centering
\includegraphics[width=\columnwidth]{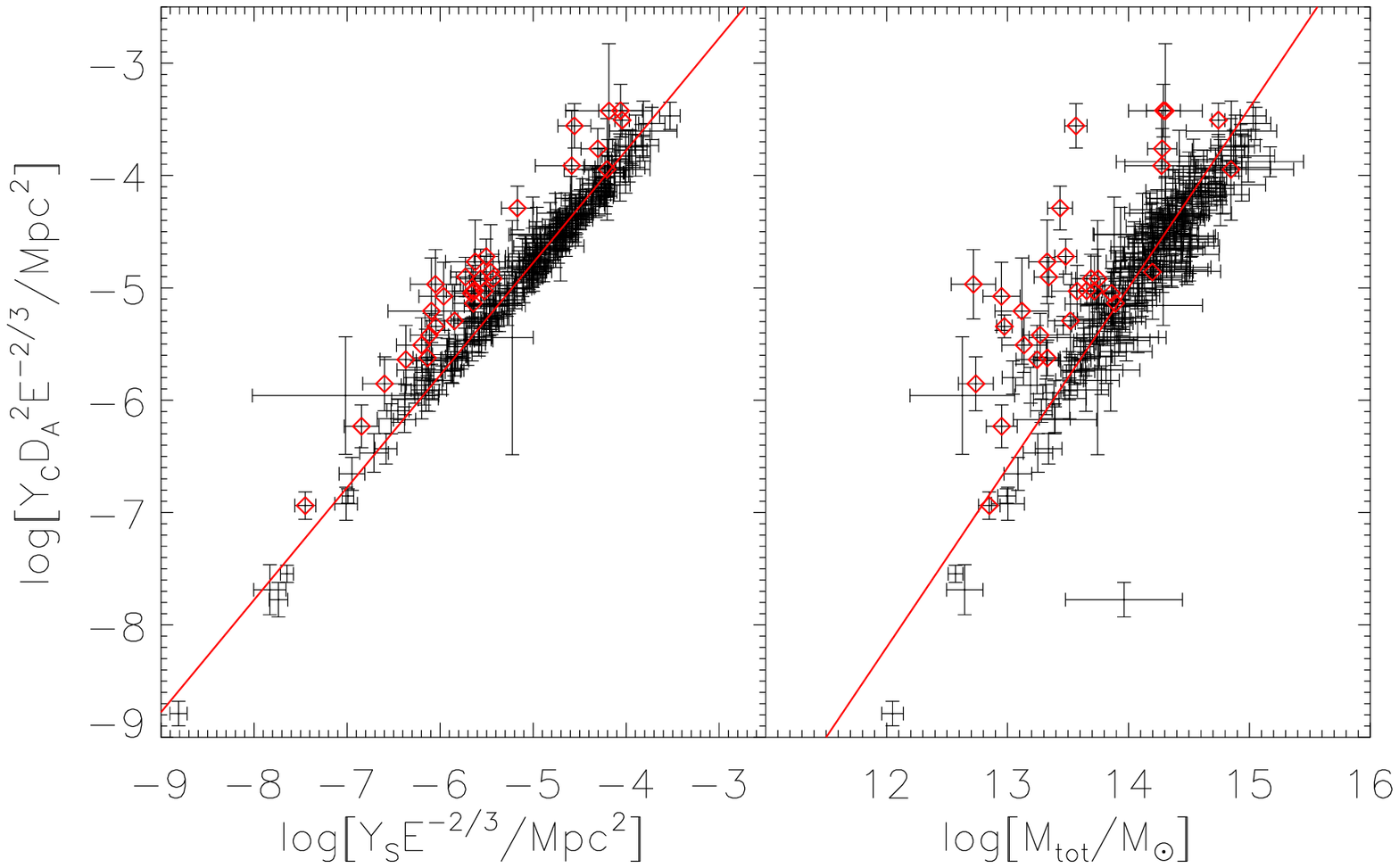}
  \caption{\itshape Left:  $Y_{C}D_{A}^{2}$ vs  $Y_{S}$,  red diamonds are clusters for which $C_{10}/C_{5}\ge 1.1$, the solid red line represents the $Y_{C}D_{A}^{2}=\bar{C}Y_{S}$, where $\bar{C}=1.67$ is the mean $C_{10}$ value, excluding the red-represented objects. Right: Logarithmic plot of $Y_{C}$ as a function of the cluster total mass $M_{tot}$, the solid red line represents the best-fitting scaling relation (Table \ref{M_Ys_param}). Clusters for which $C_{10}/C_{5}\ge 1.1$ are identified with red diamonds.}
  \label{fig_Yc_Mtot}
\end{figure}

In Table \ref{M_Ys_param_others} we compare our $Y$-$M_{tot}$ calibration with those obtained by other recent works, considering $\Delta$=2500 \citep{B2008} and $\Delta$=500 \citep{Arnaud, Andersson, PlanckPratt}. The first three lines of the table refer to observational calibrations performed by exploiting directly measured SZ fluxes (from OVRO-BIMA, SPT and Planck respectively), while the last is an X-ray only calibration, analogous to that developed in the present paper, but with a lower overdensity. In the last two columns our best-fitting normalizations are opportunely corrected for the $\Delta^{1/3}$ dependence (see eq. \ref{eq_scaling}), to match the overdensity choice of the comparison paper. The $\Delta$ dependence is in agreement with what was expected for the self-similar scenario.

\subsubsection{Redshift dependence}
In order to explore a possible redshift dependence of the considered relation, we selected nine subsamples, of 25 clusters each, sorted by $z$ (26 objects for the higher redshift interval). An iso-redshift binning would have produced an unreliable result, since our data-set does not uniformly represent the cluster population up to $z=0.89$. As shown in the upper panel of Figure \ref{fig_YsphMtot_z_bin} and by the obtained best-fitting parameters ($A=1.648\pm0.070$ and $B=-28.35\pm0.96$), no further redshift evolution, beyond the self-similar term $E(z)^{-2/3}$ already included, is evident when applying a MCMC approach with self-similar prior. By keeping the same choice of the prior for each of the nine sub-sets, the MCMC returns the coefficients shown in Figure \ref{fig_YsphMtot_z_bin} (lower panel) as a function of the mean redshifts. As displayed by the figure, the parameters are all consistent with the self-similar scaling and that obtained considering the nine average values (the black dashed line, also shown in the upper panel). The redshift independence \citep[consistent with that shown by][]{B2008} implies that our results should not be biased by the selection effect reported for the ACCEPT catalogue, where no object with $K_{0}\le 10$ $keV cm^{2}$ can be found at $z>0.1$. To strengthen and verify this result, larger redshift observations exploiting SZ and gravitational lensing measurements are needed. 

\subsubsection{$Y_{S}$ vs $Y_{C}$}
Up until now we have only considered $Y_{S}$. An X-ray estimation of $Y_{C}$ could be interesting both for investigating the goodness of the profile extrapolation (beyond $r_{2500}$, along the line of sight) and for studying scaling relations between quantities probing different volumes \citep[$M_{tot}$ spherical vs $Y$ cylindric, e.g.][]{B2008}.
$Y_{C,\Delta}$ has been calculated by considering two different truncation radii for $P_{e}(r)$, at $5r_{2500}$ (Y$_{C_{5}}$) and at $10r_{2500}$ (Y$_{C_{10}}$), both exceeding the typical cluster extent along the line of sight. 

In Figure \ref{fig_Yc_Mtot}, left panel, we compare Y$_{S}$ and $Y_{C}D_{A}^{2}$  (Y$_{C}$=Y$_{C_{10}}$). Clusters for which the ratio between the C-factors (C$_{10}$=Y$_{C}$D$_{A}^{2}$/Y$_{S}$ and C$_{5}$=Y$_{C_{5}}$D$_{A}^{2}$/Y$_{S}$, see eq. \ref{Ys} and \ref{Yc}) increases extending the truncation radius ($\frac{C_{10}}{C_{5}} > 1.1$) are marked with red diamonds. These are clusters for which the quality of X-ray data does not allow larger radii extrapolation and the un-removed background has possibly induced an offset on the observed profiles of electron density and pressure. In such cases profile extrapolation results in a diverging $Y_{C}$ value. 
If the signal to noise ratio is not high enough, profiles constrained with observations limited to small radii could result in inaccurate larger radii extrapolations, not necessarily due to the morphology of the object.

The $Y_{C}D_{A}^{2}$ - $M_{tot}$ scaling relation is reported in Figure \ref{fig_Yc_Mtot} (right panel) and the best-fitting parameters are listed in Table \ref{M_Ys_param}. Excluding the problematic clusters discussed above, we have found an average $C$ value $\bar{C}=1.67\pm0.80$. This is consistent with the results of \cite{B2008} ($\bar{C}\sim2$, with the same choice of $\Delta=2500$) even if suggesting a lower value, as we expected since we are not considering an isothermal $\beta$-model anymore. Our mean $Y_{C}/Y_{S}$ ratio is also compatible with results by \cite{Andersson} ($1.23\pm0.08$). The higher central value and the larger standard deviation are both due to the smaller projected radius considered here. Figure \ref{fig_Yc_Mtot} (right panel) and the best-fitting parameters obtained for the $Y_{C}D_{A}^{2}$-$M_{tot}$ relation (Table \ref{M_Ys_param}) confirm that the constant  $C$ assumption does not significantly affect the scaling relation calibration. 

\begin{figure}
\centering
\includegraphics[width=\columnwidth]{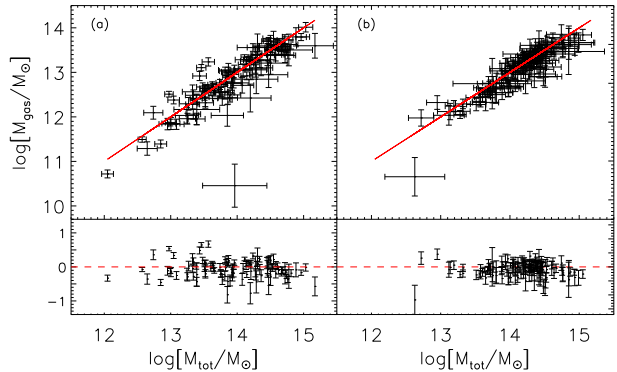}
  \caption{\itshape Comparison between $M_{gas}$ and $M_{tot}$ for the whole sample of clusters, with panels $(a)$ and $(b)$ considering the $K_{0}\le35 keV cm^{2}$ ($\bar{f}_{gas}=0.104\pm 0.074$) and the $K_{0} > 35 keV cm^{2}$ ($\bar{f}_{gas}=0.099\pm 0.037$) subsamples respectively. Solid lines correspond to $M_{gas}=f_{gas}M_{tot}$ with $f_{gas}=0.1$. At the bottom of plots $(a)$ and $(b)$, deviations from $M_{gas}=0.1 M_{tot}$ are reported, providing a graphic representation of the intrinsic scatter relevance.}\label{fig_fgas_all}
\end{figure}

\begin{figure}
 \begin{minipage}[!b]{\columnwidth}
 \centering
  \includegraphics[width=\columnwidth]{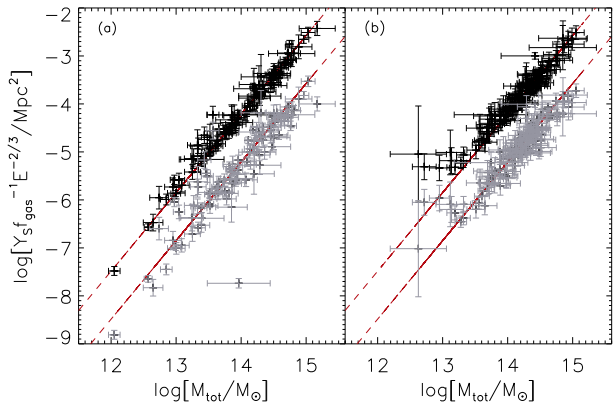}
  \caption{\itshape $Y_{S}$ vs $M_{tot}$ scaling relation with the $f_{gas}=constant$ assumption (grey) and using $f_{gas}=M_{gas}/M_{tot}$ (black). Left panel $(a)$ shows the $K_{0}\le35 keV cm^{2}$ clusters, right panel (b) the $K_{0} > 35 keV cm^{2}$ subsample.}\label{M_Y_fgas}
\end{minipage}
\end{figure}

\begin{figure}
\centering
\includegraphics[width=\columnwidth]{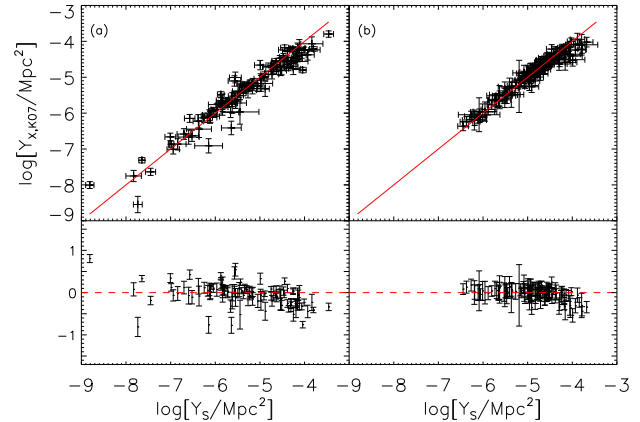}
\caption{\itshape Comparison between $Y_{X,K07}$ and $Y_{S}$.  Panel $(a)$ shows only $K_{0}\le35 keV cm^{2}$ clusters, while the left-over $K_{0} > 35 keV cm^{2}$ subsample is reported in panel $(b)$. At the bottom of plots $(a)$ and $(b)$, deviations from $Y_{X,K07}=Y_{S}$ are represented, providing a graphic representation of the intrinsic scatter relevance.}\label{fig_YkYx}
\end{figure}

\subsection{Isothermal and constant gas fraction assumptions}
Considering the $M_{gas}$-$M_{tot}$ relation (Figure \ref{fig_fgas_all}), we study the gas fraction ($f_{gas}$) behavior. The mean gas fraction over the whole sample is $\bar{f}_{gas}=0.097\pm0.057$, in agreement with past literature \citep{Vikhlinin, L2006, B2008}. At fixed overdensity, no difference between the total gas fractions obtained for CC and NCC clusters has been shown. In fact, $\bar{f}_{gas}=0.104\pm 0.074$  for the $K_{0}\le35keV cm^{2}$ subset and $\bar{f}_{gas}=0.092\pm 0.037$  for the complementary one, in agreement with that found by \cite{Chen} ($\Delta=500$) and \cite{Zhang10} ($\Delta=2500, 1000, 500$), who exploited X-ray and weak-lensing data. It is instead worth noting the larger scatter ($\sigma_{raw}=0.24$ and $\sigma_{int}=0.21$) found for the CC objects around the $M_{gas}=0.1 M_{tot}$ relation (at the bottom of Figure \ref{fig_fgas_all}). By examining this subsample, the intrinsic scatter results significantly more important (since $\sigma_{int}\sim\sigma_{raw}$) than that found for the whole cluster sample (where $\sigma_{raw}=0.21$ and $\sigma_{int}=0.12$). The assumption of $f_{gas}=const$ could then introduce an important contribution to the scatter of the $Y$-$M_{tot}$ scaling relation, especially when dealing with CC clusters (Figure \ref{M_Y_fgas}), at least at this overdensity. 

$Y$ and the $Y_{X}=M_{gas}T_{X}$ proxy \citep[$Y_{X,K07}$ hereafter, to avoid ambiguous notation, proposed by][]{K07} probe respectively the mass and the emission weighted temperatures. The comparison of these two quantities could therefore also be exploited to study the details of temperature distribution within clusters \citep[as suggested by][]{Arnaud, Andersson, PlanckPratt}. We calculate the $Y_{X,K07}$ using the mean value of the spectroscopic temperature profile observed for each cluster, without excluding the inner region since we aim at evaluating the goodness of this proxy, especially when the isothermal assumption is notoriously inappropriate. The intrinsic scatter in the $log(Y_{K,07})=A log Y_{S}+B$ relation ($A=1$ and $B=0$, Figure \ref{fig_YkYx}) again becomes significant only for the $K_{0}\le35 keV cm^{2}$ class ($\sigma_{raw}=0.26$ and $\sigma_{int}=0.19$, instead of $\sigma_{raw}=0.21$ and $\sigma_{int}=0.10$ obtained for the whole cluster sample). By constraining only $A=1$ the $Y_{K,07}/Y_{S}$ ratio can be studied. For the NCC sample the normalization obtained is consistent with zero ($B=-0.014\pm0.018$), while for the CC subset we have $B=-0.029\pm0.015$, suggesting that the $Y_{X,K07}$ proxy could underestimate the SZ signal for CCs.

\section{CC indicator}\label{CC_det}
 
\subsection{Core entropy proxy}
\begin{figure}
\centering
 \includegraphics[width=0.8\columnwidth]{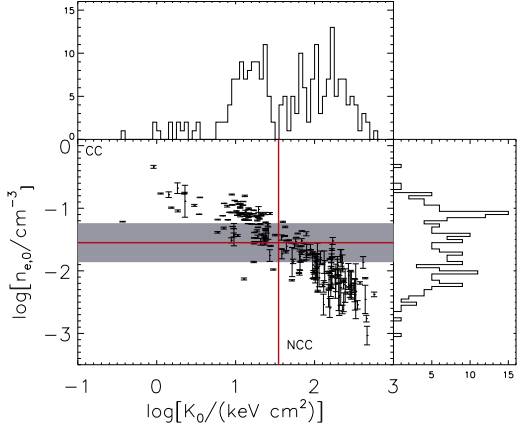}
 \caption{\itshape Correlation between $n_{e,0}$, the electron density value observed for the innermost annulus around the cluster centre, and the entropy floor $K_{0}$. The vertical red line (at $K_{0}=35 keV cm^{2}$) splits the two classes of clusters (CC and NCC) following $K_{0}$ bimodality. The grey band represents 1$\sigma$ uncertainty in the derived $n_{e,0}$ threshold, represented by the horizontal solid red line.}
\label{fig_ne0_K0}
\end{figure}

For distant objects $n_{e,0}$ is more easily determined than $t_{c,0}$ or $K_{0}$ because, even if still requiring high angular resolution, no information about the spectroscopic temperature of the electron gas is needed. For this reason $n_{e,0}$ (whose correlation with $K_{0}$ is reported in Figure \ref{fig_ne0_K0}) has already been used for the CC cluster selection in the data analysis of Planck  \citep{PlanckPratt}. However the highest significance of bimodality has been detected for $t_{c,0}$ and $K_{0}$ \citep{Cavagnolo09, Hudson}. With the central Comptonization parameter $y_{0}$ we construct a $K_{0}$ proxy that is then also independent from the X-ray spectroscopic temperature profile. The core entropy is directly related to the electron gas pressure as $K(r)=P_{e}(r)/n_{e}(r)^{5/3}$, and we propose an analogue of $K_{0}$ combining SZ and X-ray observables,
\begin{equation}
	K_{y_{0}}=\frac{y_{0}}{n_{e,0}^{5/3}}.
\end{equation}

\begin{figure}
\centering
\includegraphics[width=0.8\columnwidth]{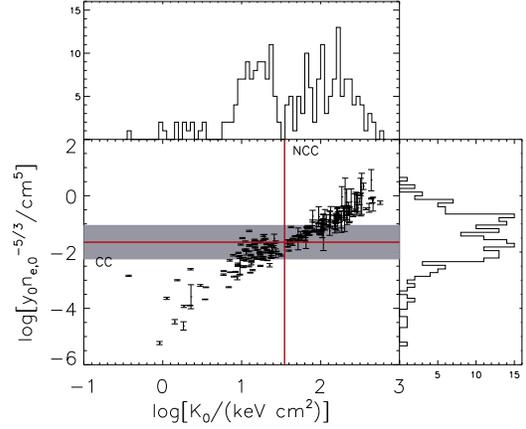}
  \caption{\itshape Entropy proxy $y_{0}/n_{e,0}^{5/3}$, constructed from the SZ central Comptonization parameter $y_{0}$, as a possible parameter for distinguishing CC from NCC clusters. The vertical red line (at $K_{0}=35 keV cm^{2}$) splits the two classes of clusters (CC and NCC) following $K_{0}$ bimodality. The grey band represents 1$\sigma$ uncertainty in the derived $y_{0}/n_{e,0}^{5/3}$ threshold, represented by the horizontal solid red line.}\label{fig_Y0_K0}
\end{figure}

Although ICM pressure is expected to preserve self-similarity better, $y_{0}$ should still probe the details of the cluster core. The $y_{0}/n_{e,0}^{5/3}$ ratio exploits the value of the innermost annulus of the observed electron density radial profile (produced by \textit{C09}), and the central Comptonization parameter calculated as the line of sight integral of our best-fitting pressure profile. Current angular resolutions of SZ experiments are in fact too low to reach direct observations of $y_{0}$. On the other hand, recent non-parametric deprojections of SZ imaging data have shown good agreement with the \textit{N07} pressure profile \citep{Basu}. The accuracy of the deprojection and the robustness of the model adopted for $P_{e}(r)$ are therefore more important than high angular resolution alone. The strong correlation found with $K_{0}$ is shown in Figure \ref{fig_Y0_K0}.
 
In Figures \ref{fig_ne0_K0} and \ref{fig_Y0_K0} we fix the CC-NCC threshold at 35 $keV cm^{2}$ once again, and convert it into a corresponding value for the two CC-indicators respectively. To do that we perform a linear fit whose returned parameters and associated uncertainties have been used to calculate the $n_{e,0}$ and $K_{y_{0}}$ thresholds, $log(n_{e,0}/cm^{-3})=-1.55\pm0.31$, $log(K_{y_{0}}/cm^{5})=-1.65\pm0.59$. Uncertainties are represented by the gray band in both plots. Clusters outside the CC and NCC regions of the plots are identified as CC according to their $K_{0}$ value, but as NCC following the proposed CC-indicator ($n_{e,0}$ or $K_{y_{0}}$), and vice versa. It is worth noting that all the objects residing in the jumbled regions of the plot fall within the gray band in Figure \ref{fig_Y0_K0}, while it is not the same in Figure \ref{fig_ne0_K0}. The $K_{y_{0}}$ histogram even shows a more pronounced bimodality. The $K_{y_{0}}$-$K_{0}$ correlation is very tight especially for NCC objects, suggesting optimistic perspectives of a useful threshold. In the context of the high quality SZ/X-ray joint analyses possible in the Planck's era, the $y_{0}n_{e,0}^{-5/3}$ ratio could then be also an interesting starting point to study the core entropy excess bypassing X-ray spectroscopy. 

\begin{figure}
\centering
\includegraphics[width=0.8\columnwidth]{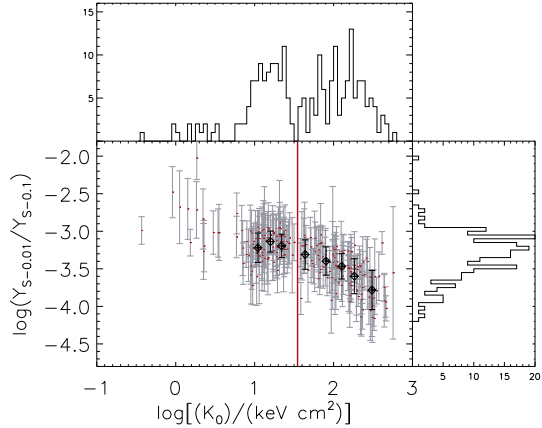}
  \caption{\itshape Correlation between the $Y_{S-0.01}/Y_{S-0.1}$ ratio and the core entropy $K_{0}$. The vertical red line (at $K_{0}=35 keV cm^{2}$) splits the two classes of clusters (CC and NCC) following $K_{0}$ bimodality. The average values of nine subsamples sorted on $K_{0}$, 25 clusters each (26 for the higher $K_{0}$ interval), are plotted in black, while gray is used to represent the single points of the whole sample.}
\label{Y_proxy}
\end{figure}

\subsection{SZ flux and $K_{0}$}
Entropy injection reduces ICM pressure at cluster cores \citep{McCarthy03b, Voit}, but its effect instead becomes negligible at larger radii, because of the dominant role of gravitational shock heating. As a consequence, the central Comptonization parameter is expected to be more sensitive than $Y$ to the entropy floor $K_{0}$. We suggest that, rather than the ratio between projected SZ signals integrated up to different angular distances from the centre \citep[probing cylindric volumes,][]{Pipino}, the same ratio calculated considering spherically integrated $Y$s, using deprojected Comptonization radial profiles, could be more sensitive to CC presence. 

In Figure \ref{Y_proxy} we represent $Y_{S-0.01}/Y_{S-0.1}$ (the ratio between the spherical Y at $0.01 r_{2500}$ and $0.1 r_{2500}$ respectively) vs $K_{0}$, in the logarithmic space. By increasing the integration radius of the quantity at denominator up to $Y_{S}$, the correlation undergoes a slight spread in comparison with that shown in Figure \ref{Y_proxy}. On the other hand the $K_{0}$ correlation disappears if we increase both the numerator and the denominator. As expected \citep{McCarthy03b, Pipino}, the $Y_{S-0.01}/Y_{S-0.1}$ ratio anti-correlates with core entropy, especially for NCC clusters, again. The correlation has been improved by considering a $K_{0}$ binning (nine subsamples, 25 clusters each, 26 for the higher $K_{0}$ interval, shown by black diamonds in Figure \ref{Y_proxy}). However, further studies exploiting SZ estimated $Y$s, instead of X-ray deduced ones, are mandatory to verify this result. In fact, this still rough diagnostic is strongly limited by the errors on $Y_{S}$, which are here dominated by uncertainties associated to the X-ray observed temperature profiles. The encouraging results of the first deprojections performed on SZ maps \citep{Basu} suggest that this correlation could be improved by exploiting SZ-observed radial pressure profiles. 

\section{Summary and conclusions}\label{conclusion}
In this paper we performed a calibration of the $Y-M_{tot}$ scaling at $\Delta=2500$ for a sample of 226 objects, observed by \textit{Chandra} and collected into the ACCEPT catalogue. Considering the entropy profiles publicly available for the ACCEPT clusters, our work exploits the core entropy $K_{0}$ in order to investigate the self-similarity of the CC and NCC populations at overdensities higher than $\Delta=500$. Our results are not very sensitive to the choice of the $K_{0}$ threshold, it can be moved from 30 $keV cm^2$ to 50 $keV cm^2$ without significantly affecting our conclusions because of the small number of objects (11) populating this core entropy region.

The universal pressure profile proposed by \cite{Nagai06} can be used to represent both CC and NCC clusters. However, when the best-fitting parameters are constrained using observations limited to relatively small radii, the analytical profiles could be affected by strong offsets (possibly due to a low signal to noise ratio and/or background removal).

The main cluster quantities calculated to develop the analysis described ($r_{2500}$, $M_{gas}$, $M_{tot}$ and $Y_{S}$) are reported in Table \ref{Tab_all}, while the main results of this paper are summarized in the list below:
\begin{itemize} 
  \item [-] The Comptonization parameter $Y$ is confirmed to be a strong mass proxy, less sensitive than other quantities to the details of cluster physics, even at $\Delta=2500$. Results of our X-ray calibration strongly agree with past literature and self-similar predictions performed at lower density contrasts. Furthermore, no redshift evolution has been identified for the best-fitting parameters obtained for the $Y$-$M_{tot}$ scaling law. However a possible additional redshift evolution cannot be completely excluded on the basis of the present result. This issue needs to be studied  further by exploiting samples extending up to larger redshifts, with an increased number of high-$z$ objects. For this purpose incoming SZ surveys are expected to provide interesting data.
  
  \item [-] Consistency with $f_{gas}=0.1$ is confirmed for both NCCs ($\bar{f}_{gas}=0.092\pm 0.037$) and CCs ($\bar{f}_{gas}=0.104\pm 0.074$), showing a wider intrinsic scatter for the latter. The constant gas fraction assumption, under which we performed our calibration, could therefore be responsible for an important contribution to the intrinsic scatter found for CCs around the best-fitting $Y-M_{tot}$ scaling.
  
  \item [-] As expected, the $Y_{X,K07}$ calculated considering the mean spectroscopic temperature of the object is a less tight $Y$ proxy for CC clusters, which are characterized by a decreasing temperature towards the center. Considering $Y_{X,K07}$ vs $Y_{S}$ in the log space and fixing the slope to unity, the normalization obtained is consistent with zero only for NCC objects. 
  \item [-]We converted the $K_{0}$ CC/NCC-threshold into a more easily observable quantity, proposed mainly for applications on higher $z$ clusters. This quantity, independent of X-ray spectroscopic data, is a core entropy proxy ($K_{y_{0}}$) showing a tighter correlation with $K_{0}$ with respect to $n_{e,0}$, when taken alone. To look for SZ-only CC indicators, we also suggest that deprojected radial Comptonization profiles should preferably be used.
\end{itemize}

\section*{Acknowledgments}
We gratefully acknowledge the referee for careful reading and useful, constructive suggestions that have significantly improved the presentation of our work. We are also thankful to Sandra Capaldi for her contribution
to the English revision of the manuscript.
Part of this work has been supported by funding from Ateneo 2010 - C26A105SWH.

\bsp

\begin{table*}
\centering
\caption{Main cluster quantities calculated in this work for the complete cluster sample of 226 objects.}
\centering
\renewcommand{\footnoterule}{}  
  \begin{tabular}{ | l c c c c c c |}
    \hline
    $cluster$ & $z$ & $D_{A}$ & $r_{2500}$ & $M_{gas}$ & $M_{tot}$ & $Y_{S}$\\
    $$ & $$ & $(Mpc)$ & $(kpc)$ & $(10^{13}M_{\odot})$ & $(10^{14}M_{\odot})$ & $(10^{-5}$ $Mpc^{2})$\\
    \hline
    \hline
$1E0657$ $56$ & $0.296$ & $910$ & $601\pm68$ & $6.0\pm1.3$ & $4.2\pm1.0$ & $9.5\pm2.0$\\
$2A$ $0335+096$ & $0.035$ & $142$ & $395\pm35$ & $0.90\pm 0.16$ & $0.89\pm 0.16$ & $0.425\pm0.060$\\
$2PIGG$ $J0011.5-2850$ & $0.075$ & $295$ & $373\pm58$ & $0.58\pm 0.23$ & $ 0.76\pm 0.20$ & $0.45\pm 0.16$\\
$2PIGG$ $J2227.0-3041$ & $0.073$ & $286$ & $312\pm47$ & $0.42\pm 0.12$ & $ 0.44\pm 0.12$ & $0.145\pm0.049$\\
$3C$ $28.0$ & $0.195$ & $668$ & $470\pm138$ & $1.202\pm 0.611$ & $1.61\pm 0.90$ & $1.05\pm 0.64$\\
$3C$ $295$ & $0.464$ & $ 1209$ & $634\pm94$ & $2.47\pm 0.51$ & $5.8\pm1.2$ & $4.12\pm 0.94$\\
$3C$ $388$ & $0.092$ & $352$ & $347\pm61$ & $0.256\pm0.092$ & $ 0.60\pm 0.20$ & $0.108\pm0.031$\\
$4C$ $55.16$ & $0.242$ & $787$ & $429\pm54$ & $0.84\pm 0.20$ & $1.40\pm 0.28$ & $0.59\pm 0.13$\\
$Abell$ $13$ & $0.094$ & $360$ & $468\pm108$ & $0.90\pm 0.48$ & $1.45\pm 0.47$ & $0.87\pm 0.45$\\
$Abell$ $68$ & $0.255$ & $817$ & $547\pm112$ & $2.6\pm1.1$ & $2.8\pm1.3$ & $3.5\pm1.1$\\
$Abell$ $85$ & $0.056$ & $223$ & $443\pm29$ & $2.20\pm 0.30$ & $1.29\pm 0.16$ & $1.78\pm 0.28$\\
$Abell$ $119$ & $0.044$ & $179$ & $213\pm64$ & $0.11\pm 0.12$ & $ 0.131\pm0.067$ & $0.082\pm0.088$\\
$Abell$ $133$ & $0.056$ & $223$ & $365\pm31$ & $0.83\pm 0.14$ & $ 0.71\pm 0.11$ & $0.389\pm0.073$\\
$Abell$ $141$ & $0.230$ & $758$ & $712\pm176$ & $3.4\pm1.7$ & $5.8\pm2.8$ & $6.9\pm4.2$\\
$Abell$ $160$ & $0.045$ & $181$ & $236\pm25$ & $0.113\pm0.033$ & $ 0.190\pm0.039$ & $0.0357\pm0.0098$\\
$Abell$ $209$ & $0.206$ & $696$ & $461\pm56$ & $1.84\pm 0.51$ & $1.65\pm 0.38$ & $1.95\pm 0.58$\\
$Abell$ $222$ & $0.213$ & $715$ & $326\pm53$ & $0.57\pm 0.22$ & $ 0.57\pm 0.20$ & $0.34\pm 0.13$\\
$Abell$ $223$ & $0.207$ & $699$ & $389\pm96$ & $0.43\pm 0.23$ & $ 0.94\pm 0.40$ & $0.34\pm 0.19$\\
$Abell$ $262$ & $0.016$ & $69$ & $199\pm24$ & $0.239\pm0.071$ & $ 0.111\pm0.022$ & $0.057\pm0.018$\\
$Abell$ $267$ & $0.230$ & $758$ & $514\pm99$ & $1.57\pm 0.51$ & $2.5\pm1.3$ & $1.61\pm 0.53$\\
$Abell$ $368$ & $0.220$ & $733$ & $419\pm72$ & $1.01\pm 0.33$ & $1.18\pm 0.39$ & $0.68\pm 0.29$\\
$Abell$ $370$ & $0.375$ & $ 1064$ & $745\pm81$ & $5.9\pm1.5$ & $8.5\pm1.7$ & $16.6\pm3.6$\\
$Abell$ $383$ & $0.187$ & $646$ & $438\pm47$ & $1.79\pm 0.34$ & $1.39\pm 0.30$ & $1.23\pm 0.27$\\
$Abell$ $399$ & $0.072$ & $281$ & $454\pm69$ & $1.28\pm 0.41$ & $1.39\pm 0.45$ & $1.21\pm 0.41$\\
$Abell$ $401$ & $0.075$ & $292$ & $596\pm143$ & $2.5\pm1.1$ & $2.9\pm1.3$ & $2.9\pm1.3$\\
$Abell$ $478$ & $0.088$ & $340$ & $570\pm80$ & $3.65\pm 0.83$ & $2.8\pm1.1$ & $3.38\pm 0.69$\\
$Abell$ $496$ & $0.033$ & $135$ & $358\pm37$ & $1.07\pm 0.23$ & $ 0.66\pm 0.13$ & $0.50\pm 0.11$\\
$Abell$ $520$ & $0.202$ & $686$ & $379\pm91$ & $1.21\pm 0.79$ & $ 0.86\pm 0.45$ & $1.17\pm 0.75$\\
$Abell$ $521$ & $0.253$ & $814$ & $320\pm68$ & $0.32\pm 0.16$ & $ 0.56\pm 0.24$ & $0.30\pm 0.18$\\
$Abell$ $539$ & $0.029$ & $119$ & $246\pm34$ & $0.191\pm0.065$ & $ 0.214\pm0.051$ & $0.074\pm0.026$\\
$Abell$ $562$ & $0.110$ & $414$ & $239\pm34$ & $0.196\pm0.085$ & $ 0.209\pm0.052$ & $0.084\pm0.032$\\
$Abell$ $586$ & $0.171$ & $601$ & $472\pm53$ & $2.09\pm 0.50$ & $1.70\pm 0.42$ & $1.76\pm 0.37$\\
$Abell$ $611$ & $0.288$ & $893$ & $495\pm112$ & $0.92\pm 0.35$ & $2.3\pm1.6$ & $1.05\pm 0.38$\\
$Abell$ $644$ & $0.070$ & $275$ & $573\pm67$ & $2.74\pm 0.60$ & $2.78\pm 0.61$ & $2.93\pm 0.76$\\
$Abell$ $665$ & $0.181$ & $629$ & $433\pm82$ & $1.77\pm 0.75$ & $1.31\pm 0.44$ & $1.86\pm 0.87$\\
$Abell$ $697$ & $0.282$ & $880$ & $512\pm108$ & $3.4\pm1.4$ & $2.3\pm1.2$ & $4.4\pm2.0$\\
$Abell$ $744$ & $0.073$ & $286$ & $283\pm45$ & $0.189\pm0.075$ & $ 0.33\pm 0.10$ & $0.065\pm0.027$\\
$Abell$ $773$ & $0.217$ & $725$ & $562\pm115$ & $2.8\pm1.0$ & $2.9\pm1.1$ & $3.1\pm1.3$\\
$Abell$ $907$ & $0.153$ & $547$ & $531\pm116$ & $1.26\pm 0.41$ & $2.29\pm 0.92$ & $1.04\pm 0.35$\\
$Abell$ $963$ & $0.206$ & $695$ & $491\pm57$ & $1.95\pm 0.49$ & $2.00\pm 0.42$ & $1.91\pm 0.43$\\
$Abell$ $1063S$ & $0.354$ & $ 1026$ & $662\pm163$ & $7.1\pm2.8$ & $5.5\pm2.7$ & $13.0\pm5.0$\\
$Abell$ $1068$ & $0.138$ & $501$ & $596\pm59$ & $3.34\pm 0.61$ & $3.34\pm 0.64$ & $4.40\pm 0.97$\\
$Abell$ $1201$ & $0.169$ & $594$ & $316\pm68$ & $0.56\pm 0.31$ & $ 0.49\pm 0.18$ & $0.39\pm 0.23$\\
$Abell$ $1204$ & $0.171$ & $599$ & $356\pm42$ & $1.03\pm 0.20$ & $ 0.73\pm 0.17$ & $0.46\pm 0.11$\\
$Abell$ $1240$ & $0.159$ & $566$ & $157\pm55$ & $0.005\pm0.048$ & $0.042\pm0.042$ & $0.010\pm0.023$\\
$Abell$ $1361$ & $0.117$ & $437$ & $297\pm41$ & $0.39\pm 0.12$ & $ 0.40\pm 0.11$ & $0.135\pm0.033$\\
$Abell$ $1413$ & $0.143$ & $517$ & $465\pm46$ & $2.39\pm 0.50$ & $1.60\pm 0.35$ & $2.02\pm 0.51$\\
$Abell$ $1423$ & $0.213$ & $715$ & $426\pm103$ & $1.09\pm 0.53$ & $1.21\pm 0.57$ & $0.83\pm 0.48$\\
$Abell$ $1446$ & $0.104$ & $392$ & $314\pm44$ & $0.39\pm 0.14$ & $ 0.47\pm 0.11$ & $0.225\pm0.077$\\
$Abell$ $1569$ & $0.074$ & $288$ & $319\pm67$ & $0.211\pm0.085$ & $ 0.45\pm 0.19$ & $0.078\pm0.034$\\
$Abell$ $1576$ & $0.279$ & $873$ & $847\pm186$ & $4.7\pm1.8$ & $9.8\pm4.9$ & $11.7\pm6.2$\\
$Abell$ $1644$ & $0.047$ & $191$ & $248\pm33$ & $0.306\pm0.098$ & $ 0.219\pm0.061$ & $0.190\pm0.070$\\
$Abell$ $1650$ & $0.084$ & $327$ & $397\pm54$ & $1.61\pm 0.50$ & $ 0.93\pm 0.24$ & $1.15\pm 0.36$\\
$Abell$ $1651$ & $0.084$ & $325$ & $598\pm167$ & $1.86\pm 0.81$ & $2.77\pm1.69$ & $1.90\pm 0.82$\\
$Abell$ $1664$ & $0.128$ & $470$ & $506\pm92$ & $1.14\pm 0.41$ & $1.91\pm 0.71$ & $0.90\pm 0.24$\\
$Abell$ $1689$ & $0.184$ & $638$ & $627\pm89$ & $4.6\pm1.1$ & $4.0\pm1.1$ & $5.9\pm1.4$\\
$Abell$ $1736$ & $0.034$ & $139$ & $217\pm46$ & $0.132\pm0.085$ & $ 0.137\pm0.059$ & $0.064\pm0.040$\\
$Abell$ $1758$ & $0.279$ & $874$ & $144\pm27$ & $0.095\pm0.049$ & $0.052\pm0.022$ & $0.098\pm0.061$\\
$Abell$ $1763$ & $0.187$ & $644$ & $521\pm87$ & $2.29\pm 0.80$ & $2.29\pm 0.64$ & $2.61\pm 0.86$\\
$Abell$ $1795$ & $0.063$ & $248$ & $644\pm107$ & $2.48\pm 0.49$ & $3.8\pm1.2$ & $2.31\pm 0.51$\\
$Abell$ $1835$ & $0.253$ & $814$ & $591\pm100$ & $5.8\pm1.6$ & $3.5\pm1.3$ & $7.2\pm2.7$\\
$Abell$ $1914$ & $0.171$ & $601$ & $683\pm138$ & $3.9\pm1.1$ & $5.8\pm3.1$ & $5.7\pm1.3$\\
\hline
  \end{tabular}
\end{table*}

\begin{table*}
\centering
\setcounter{table}{2}
\caption{continued.}
\renewcommand{\footnoterule}{}  
  \begin{tabular}{ | l c c c c c c |}
    \hline
    $cluster$ & $z$ & $D_{A}$ & $r_{2500}$ & $M_{gas}$ & $M_{tot}$ & $Y_{S}$\\
    $$ & $$ & $(Mpc)$ & $(kpc)$ & $(10^{13}M_{\odot})$ & $(10^{14}M_{\odot})$ & $(10^{-5}$ $Mpc^{2})$\\
    \hline
    \hline
$Abell$ $1942$ & $0.224$ & $743$ & $365\pm117$ & $0.45\pm 0.37$ & $ 0.63\pm 0.48$ & $0.43\pm 0.28$\\
$Abell$ $1991$ & $0.059$ & $234$ & $279\pm22$ & $0.431\pm0.077$ & $ 0.323\pm0.052$ & $0.133\pm0.023$\\
$Abell$ $1995$ & $0.319$ & $957$ & $489\pm92$ & $1.27\pm 0.47$ & $2.07\pm 0.95$ & $1.33\pm 0.44$\\
$Abell$ $2029$ & $0.077$ & $299$ & $455\pm109$ & $1.9\pm1.1$ & $1.52\pm 0.16$ & $2.91\pm 0.52$\\
$Abell$ $2034$ & $0.113$ & $424$ & $498\pm57$ & $1.73\pm 0.43$ & $1.94\pm 0.37$ & $1.73\pm 0.41$\\
$Abell$ $2052$ & $0.035$ & $145$ & $307\pm24$ & $0.526\pm0.082$ & $ 0.421\pm0.064$ & $0.174\pm0.029$\\
$Abell$ $2063$ & $0.035$ & $144$ & $395\pm66$ & $0.69\pm 0.20$ & $ 0.90\pm 0.32$ & $0.358\pm0.093$\\
$Abell$ $2065$ & $0.073$ & $287$ & $375\pm39$ & $0.92\pm 0.24$ & $ 0.79\pm 0.15$ & $0.72\pm 0.17$\\
$Abell$ $2069$ & $0.116$ & $433$ & $295\pm66$ & $0.27\pm 0.18$ & $ 0.38\pm 0.18$ & $0.23\pm 0.15$\\
$Abell$ $2104$ & $0.155$ & $555$ & $463\pm54$ & $2.17\pm 0.60$ & $1.55\pm 0.39$ & $2.03\pm 0.61$\\
$Abell$ $2107$ & $0.041$ & $167$ & $316\pm34$ & $0.356\pm0.084$ & $ 0.452\pm0.086$ & $0.238\pm0.067$\\
$Abell$ $2111$ & $0.230$ & $758$ & $406\pm91$ & $1.22\pm 0.66$ & $1.12\pm 0.50$ & $1.15\pm 0.62$\\
$Abell$ $2124$ & $0.066$ & $260$ & $322\pm48$ & $0.37\pm 0.15$ & $ 0.48\pm 0.12$ & $0.185\pm0.069$\\
$Abell$ $2125$ & $0.247$ & $798$ & $285\pm42$ & $0.277\pm0.099$ & $ 0.40\pm 0.11$ & $0.130\pm0.047$\\
$Abell$ $2142$ & $0.090$ & $346$ & $528\pm36$ & $3.43\pm 0.49$ & $2.25\pm 0.26$ & $3.85\pm 0.58$\\
$Abell$ $2147$ & $0.036$ & $146$ & $343\pm69$ & $0.35\pm 0.15$ & $ 0.63\pm 0.35$ & $0.221\pm0.099$\\
$Abell$ $2151$ & $0.037$ & $150$ & $237\pm38$ & $0.197\pm0.075$ & $ 0.186\pm0.061$ & $0.078\pm0.037$\\
$Abell$ $2163$ & $0.170$ & $596$ & $830\pm323$ & $4.6\pm3.1$ & $7.1\pm6.1$ & $16\pm13$\\
$Abell$ $2187$ & $0.183$ & $635$ & $753\pm361$ & $2.94\pm 0.49$ & $7.0\pm8.4$ & $6.6\pm7.0$\\
$Abell$ $2199$ & $0.030$ & $124$ & $266\pm37$ & $0.83\pm 0.29$ & $ 0.272\pm0.064$ & $0.68\pm0.27$\\
$Abell$ $2204$ & $0.152$ & $546$ & $830\pm79$ & $6.20\pm 0.92$ & $9.2\pm1.7$ & $15.8\pm3.0$\\
$Abell$ $2218$ & $0.171$ & $601$ & $534\pm82$ & $2.37\pm 0.65$ & $2.39\pm 0.73$ & $2.19\pm 0.69$\\
$Abell$ $2219$ & $0.226$ & $747$ & $585\pm80$ & $5.0\pm1.5$ & $3.44\pm 0.91$ & $8.3\pm2.8$\\
$Abell$ $2244$ & $0.097$ & $369$ & $419\pm50$ & $1.73\pm 0.47$ & $1.11\pm 0.23$ & $1.23\pm 0.32$\\
$Abell$ $2256$ & $0.058$ & $231$ & $509\pm76$ & $2.39\pm 0.97$ & $1.90\pm 0.53$ & $5.1\pm2.1$\\
$Abell$ $2259$ & $0.164$ & $581$ & $476\pm110$ & $1.58\pm 0.65$ & $1.64\pm 0.64$ & $1.17\pm 0.53$\\
$Abell$ $2261$ & $0.224$ & $743$ & $553\pm126$ & $2.9\pm1.1$ & $2.7\pm1.2$ & $3.0\pm1.3$\\
$Abell$ $2294$ & $0.178$ & $620$ & $510\pm56$ & $1.97\pm 0.53$ & $2.16\pm 0.44$ & $2.47\pm0.63$\\
$Abell$ $2384$ & $0.095$ & $362$ & $370\pm38$ & $0.66\pm 0.16$ & $ 0.76\pm 0.12$ & $0.344\pm0.060$\\
$Abell$ $2390$ & $0.230$ & $758$ & $573\pm89$ & $5.0\pm1.6$ & $3.17\pm 0.96$ & $7.0\pm2.7$\\
$Abell$ $2409$ & $0.148$ & $533$ & $414\pm67$ & $1.41\pm 0.51$ & $1.11\pm 0.27$ & $1.08\pm 0.39$\\
$Abell$ $2420$ & $0.085$ & $328$ & $676\pm139$ & $2.9\pm1.4$ & $4.3\pm1.6$ & $5.2\pm2.4$\\
$Abell$ $2462$ & $0.074$ & $289$ & $289\pm37$ & $0.223\pm0.068$ & $ 0.359\pm0.083$ & $0.087\pm0.025$\\
$Abell$ $2537$ & $0.295$ & $908$ & $468\pm53$ & $2.33\pm 0.58$ & $1.92\pm 0.41$ & $2.52\pm 0.65$\\
$Abell$ $2554$ & $0.110$ & $415$ & $446\pm84$ & $0.64\pm 0.27$ & $1.29\pm 0.43$ & $0.54\pm 0.23$\\
$Abell$ $2556$ & $0.086$ & $333$ & $338\pm65$ & $0.51\pm 0.17$ & $1.19\pm 0.48$ & $0.269\pm0.088$\\
$Abell$ $2589$ & $0.042$ & $169$ & $184\pm36$ & $0.21\pm 0.13$ & $0.089\pm0.032$ & $0.111\pm0.068$\\
$Abell$ $2597$ & $0.085$ & $330$ & $268\pm21$ & $1.27\pm 0.22$ & $ 0.288\pm0.047$ & $0.366\pm0.065$\\
$Abell$ $2626$ & $0.057$ & $229$ & $308\pm33$ & $0.47\pm 0.12$ & $ 0.430\pm0.084$ & $0.189\pm0.044$\\
$Abell$ $2631$ & $0.278$ & $871$ & $515\pm62$ & $2.82\pm 0.64$ & $2.46\pm 0.56$ & $2.92\pm 0.72$\\
$Abell$ $2657$ & $0.040$ & $164$ & $334\pm79$ & $0.44\pm 0.25$ & $ 0.50\pm 0.18$ & $0.24\pm 0.13$\\
$Abell$ $2667$ & $0.230$ & $758$ & $463\pm46$ & $3.62\pm 0.74$ & $1.73\pm 0.34$ & $3.49\pm 0.85$\\
$Abell$ $2717$ & $0.048$ & $192$ & $286\pm46$ & $0.208\pm0.078$ & $ 0.332\pm0.094$ & $0.076\pm0.027$\\
$Abell$ $2744$ & $0.308$ & $936$ & $850\pm134$ & $6.9\pm2.1$ & $11.3\pm3.3$ & $21.0\pm6.8$\\
$Abell$ $2813$ & $0.292$ & $903$ & $542\pm142$ & $1.17\pm 0.59$ & $2.7\pm1.2$ & $1.51\pm 0.73$\\
$Abell$ $3084$ & $0.098$ & $373$ & $377\pm71$ & $0.35\pm 0.15$ & $ 0.77\pm 0.28$ & $0.203\pm0.091$\\
$Abell$ $3088$ & $0.253$ & $815$ & $578\pm131$ & $1.32\pm 0.57$ & $3.3\pm1.7$ & $1.94\pm 0.75$\\
$Abell$ $3112$ & $0.072$ & $283$ & $392\pm47$ & $1.20\pm 0.30$ & $ 0.91\pm 0.18$ & $0.69\pm 0.16$\\
$Abell$ $3120$ & $0.069$ & $272$ & $245\pm29$ & $0.115\pm0.029$ & $ 0.218\pm0.056$ & $0.0267\pm0.0073$\\
$Abell$ $3158$ & $0.058$ & $232$ & $389\pm32$ & $1.06\pm 0.19$ & $ 0.86\pm 0.12$ & $0.74\pm 0.15$\\
$Abell$ $3266$ & $0.059$ & $235$ & $380\pm75$ & $1.08\pm 0.56$ & $ 0.76\pm 0.29$ & $1.01\pm 0.60$\\
$Abell$ $3364$ & $0.148$ & $534$ & $870\pm265$ & $4.2\pm2.2$ & $8.6\pm4.8$ & $9.0\pm5.6$\\
$Abell$ $3528S$ & $0.053$ & $213$ & $398\pm72$ & $0.58\pm 0.22$ & $ 0.88\pm0.23$ & $0.34\pm 0.12$\\
$Abell$ $3571$ & $0.039$ & $160$ & $519\pm102$ & $4.0\pm2.1$ & $1.95\pm 0.63$ & $8.8\pm4.7$\\
$Abell$ $3581$ & $0.022$ & $91$ & $259\pm34$ & $0.255\pm0.075$ & $ 0.242\pm0.051$ & $0.069\pm0.017$\\
$Abell$ $3667$ & $0.056$ & $223$ & $408\pm54$ & $1.12\pm 0.35$ & $1.02\pm 0.18$ & $0.80\pm 0.22$\\
$Abell$ $3822$ & $0.076$ & $297$ & $337\pm95$ & $0.42\pm 0.35$ & $ 0.52\pm 0.24$ & $0.31\pm 0.24$\\
$Abell$ $3827$ & $0.098$ & $375$ & $563\pm428$ & $2.58\pm 0.95$ & $2.6\pm3.1$ & $10.2\pm1.6$\\
$Abell$ $3921$ & $0.093$ & $356$ & $420\pm84$ & $1.02\pm 0.46$ & $1.08\pm 0.37$ & $0.84\pm 0.41$\\
$Abell$ $4038$ & $0.030$ & $124$ & $411\pm97$ & $0.47\pm 0.17$ & $ 0.96\pm 0.47$ & $0.222\pm0.051$\\
$Abell$ $4059$ & $0.048$ & $192$ & $275\pm31$ & $0.47\pm 0.12$ & $ 0.303\pm0.064$ & $0.319\pm0.096$\\
$Abell$ $S0405$ & $0.061$ & $244$ & $289\pm73$ & $0.26\pm 0.18$ & $ 0.33\pm 0.14$ & $0.15\pm 0.11$\\
\hline
  \end{tabular}
\end{table*}

\begin{table*}
\centering
\setcounter{table}{2}
\caption{continued.}
\renewcommand{\footnoterule}{}  
  \begin{tabular}{ | l c c c c c c |}
    \hline
    $cluster$ & $z$ & $D_{A}$ & $r_{2500}$ & $M_{gas}$ & $M_{tot}$ & $Y_{S}$\\
    $$ & $$ & $(Mpc)$ & $(kpc)$ & $(10^{13}M_{\odot})$ & $(10^{14}M_{\odot})$ & $(10^{-5}$ $Mpc^{2})$\\
    \hline
    \hline
$Abell$ $S0592$ & $0.222$ & $737$ & $493\pm82$ & $2.01\pm 0.64$ & $2.03\pm 0.75$ & $2.7\pm1.0$\\
$AC$ $114$ & $0.312$ & $944$ & $436\pm72$ & $1.58\pm 0.56$ & $1.56\pm 0.45$ & $1.65\pm 0.62$\\
$AWM7$ & $0.017$ & $72$ & $245\pm60$ & $0.37\pm 0.24$ & $ 0.213\pm0.075$ & $0.24\pm 0.18$\\
$CENTAURUS$ & $0.011$ & $46$ & $727\pm68$ & $2.84\pm 0.50$ & $5.54\pm 0.68$ & $9.1\pm1.6$\\
$CID$ $0072$ & $0.034$ & $141$ & $235\pm40$ & $0.147\pm0.062$ & $ 0.25\pm 0.20$ & $0.0419\pm0.012$\\
$CL$ $J1226.9+3332$ & $0.89$ & $ 1602$ & $452\pm67$ & $2.81\pm 0.71$ & $3.38\pm 0.71$ & $4.3\pm1.1$\\
$CYGNUS A$ & $0.056$ & $225$ & $461\pm79$ & $1.50\pm 0.45$ & $1.36\pm 0.37$ & $1.02\pm 0.30$\\
$ESO$ $3060170$ & $0.036$ & $147$ & $371\pm63$ & $0.269\pm0.091$ & $ 0.72\pm 0.24$ & $0.220\pm0.093$\\
$ESO$ $5520200$ & $0.031$ & $129$ & $233\pm52$ & $0.108\pm0.058$ & $ 0.176\pm0.071$ & $0.043\pm0.028$\\
$EXO$ $0422-086$ & $0.040$ & $162$ & $286\pm37$ & $0.44\pm 0.12$ & $ 0.336\pm0.067$ & $0.135\pm0.033$\\
$HCG$ $0062$ & $0.015$ & $61$ & $138\pm11$ & $0.0311\pm0.0053$ & $0.0372\pm0.0052$ & $0.00227\pm 0.00036$\\
$HCG$ $42$ & $0.013$ & $56$ & $430\pm255$ & $0.0029\pm0.0015$ & $ 0.9\pm1.0$ & $0.00184\pm 0.00042$\\
$HERCULES A$ & $0.154$ & $551$ & $423\pm63$ & $1.12\pm 0.29$ & $1.23\pm 0.40$ & $0.66\pm 0.15$\\
$HYDRA$ $A$ & $0.055$ & $220$ & $361\pm37$ & $0.98\pm 0.19$ & $ 0.6836\pm0.094$ & $0.456\pm0.091$\\
$M49$ & $ 0.003$ & $14$ & $93\pm11$ & $0.0052\pm0.0011$ & $0.0112\pm0.0023$ & $0.000154\pm0.000033$\\
$M87$ & $ 0.004$ & $19$ & $188\pm16$ & $0.324\pm0.074$ & $0.094\pm0.014$ & $0.090\pm0.020$\\
$MACS$ $J0011.7-1523$ & $0.360$ & $ 1038$ & $560\pm97$ & $3.2\pm1.1$ & $3.38\pm 0.89$ & $4.9\pm1.6$\\
$MACS$ $J0035.4-2015$ & $0.364$ & $ 1046$ & $521\pm95$ & $3.16\pm1.0$ & $2.83\pm 0.84$ & $3.5\pm1.1$\\
$MACS$ $J0159.8-0849$ & $0.405$ & $ 1117$ & $745\pm176$ & $6.28\pm2.8$ & $7.7\pm3.6$ & $14.6\pm8.3$\\
$MACS$ $J0242.5-2132$ & $0.314$ & $948$ & $654\pm61$ & $3.78\pm 0.66$ & $5.36\pm 0.89$ & $6.41\pm 0.96$\\
$MACS$ $J0257.1-2325$ & $0.505$ & $ 1266$ & $517\pm108$ & $0.79\pm 0.35$ & $3.0\pm1.4$ & $1.23\pm 0.44$\\
$MACS$ $J0257.6-2209$ & $0.322$ & $965$ & $526\pm90$ & $1.78\pm 0.53$ & $2.71\pm 0.88$ & $1.89\pm 0.54$\\
$MACS$ $J0308.9+2645$ & $0.324$ & $968$ & $628\pm85$ & $4.2\pm1.1$ & $4.6\pm1.3$ & $6.6\pm1.7$\\
$MACS$ $J0329.6-0211$ & $0.450$ & $ 1188$ & $664\pm74$ & $4.23\pm 0.79$ & $6.5\pm1.5$ & $8.7\pm2.0$\\
$MACS$ $J0417.5-1154$ & $0.440$ & $ 1173$ & $473\pm149$ & $3.6\pm2.4$ & $2.0\pm1.4$ & $7.6\pm8.2$\\
$MACS$ $J0429.6-0253$ & $0.399$ & $ 1107$ & $604\pm63$ & $3.68\pm 0.75$ & $4.6\pm1.0$ & $6.3\pm1.2$\\
$MACS$ $J0520.7-1328$ & $0.340$ & $999$ & $503\pm64$ & $2.41\pm 0.58$ & $2.46\pm 0.68$ & $2.38\pm 0.56$\\
$MACS$ $J0547.0-3904$ & $0.210$ & $707$ & $339\pm53$ & $0.349\pm0.098$ & $ 0.65\pm 0.21$ & $0.146\pm0.041$\\
$MACS$ $J0717.5+3745$ & $0.548$ & $ 1320$ & $507\pm67$ & $4.3\pm1.2$ & $3.24\pm 0.72$ & $7.2\pm1.8$\\
$MACS$ $J0744.8+3927$ & $0.686$ & $ 1462$ & $639\pm90$ & $5.3\pm1.4$ & $7.4\pm2.1$ & $14.4\pm4.3$\\
$MACS$ $J1115.2+5320$ & $0.439$ & $ 1171$ & $453\pm66$ & $2.16\pm 0.68$ & $2.01\pm 0.57$ & $2.34\pm 0.69$\\
$MACS$ $J1115.8+0129$ & $0.352$ & $ 1023$ & $528\pm75$ & $3.93\pm 0.89$ & $2.94\pm 0.93$ & $4.1\pm1.0$\\
$MACS$ $J1131.8-1955$ & $0.3070$ & $933$ & $448\pm91$ & $1.52\pm 0.67$ & $1.56\pm 0.77$ & $1.70\pm 0.89$\\
$MACS$ $J1149.5+2223$ & $0.544$ & $ 1315$ & $397\pm63$ & $2.2\pm 0.74$ & $1.50\pm 0.57$ & $2.7\pm 1.0$\\
$MACS$ $J1206.2-0847$ & $0.440$ & $ 1173$ & $511\pm86$ & $4.1\pm1.3$ & $2.84\pm 0.97$ & $5.6\pm2.0$\\
$MACS$ $J1311.0-0310$ & $0.494$ & $ 1251$ & $429\pm56$ & $1.67\pm 0.32$ & $1.84\pm 0.49$ & $1.29\pm 0.33$\\
$MACS$ $J1621.3+3810$ & $0.461$ & $ 1204$ & $889\pm242$ & $4.0\pm1.3$ & $14.8\pm9.5$ & $11.8\pm4.0$\\
$MACS$ $J1931.8-2634$ & $0.352$ & $ 1023$ & $645\pm64$ & $5.3\pm1.0$ & $5.2\pm1.4$ & $9.8\pm2.3$\\
$MACS$ $J2049.9-3217$ & $0.325$ & $971$ & $502\pm118$ & $2.05\pm 0.99$ & $2.2\pm1.1$ & $2.46\pm 0.82$\\
$MACS$ $J2211.7-0349$ & $0.270$ & $853$ & $687\pm101$ & $4.23\pm 0.99$ & $5.7\pm1.6$ & $6.5\pm1.5$\\
$MACS$ $J2214.9-1359$ & $0.503$ & $ 1263$ & $535\pm130$ & $0.81\pm 0.43$ & $3.2\pm1.8$ & $1.47\pm0.63$\\
$MACS$ $J2228+2036$ & $0.412$ & $ 1128$ & $469\pm72$ & $2.53\pm 0.84$ & $2.15\pm 0.70$ & $3.01\pm 0.95$\\
$MACS$ $J2229.7-2755$ & $0.324$ & $968$ & $619\pm59$ & $3.12\pm 0.56$ & $4.6\pm1.0$ & $4.84\pm 0.88$\\
$MACS$ $J2245.0+2637$ & $0.304$ & $927$ & $451\pm97$ & $1.36\pm 0.45$ & $1.57\pm 0.98$ & $1.01\pm 0.33$\\
$MKW 04$ & $0.020$ & $83$ & $192\pm25$ & $0.071\pm0.017$ & $ 0.101\pm0.031$ & $0.0098\pm0.0028$\\
$MKW 08$ & $0.027$ & $112$ & $223\pm25$ & $0.174\pm0.057$ & $ 0.155\pm0.045$ & $0.054\pm0.018$\\
$MKW3S$ & $0.045$ & $183$ & $357\pm34$ & $0.63\pm 0.12$ & $ 0.67\pm 0.14$ & $0.270\pm0.044$\\
$MS$ $0016.9+1609$ & $0.541$ & $ 1312$ & $474\pm129$ & $2.9\pm1.6$ & $2.5\pm1.2$ & $4.2\pm2.2$\\
$MS$ $0116.3-0115$ & $0.045$ & $183$ & $258\pm39$ & $0.090\pm0.033$ & $ 0.245\pm0.072$ & $0.030\pm0.010$\\
$MS$ $0440.5+0204$ & $0.190$ & $654$ & $690\pm118$ & $2.40\pm 0.88$ & $5.3\pm1.9$ & $5.6\pm2.2$\\
$MS$ $0451.6-0305$ & $0.539$ & $ 1309$ & $431\pm119$ & $0.64\pm 0.39$ & $1.7\pm1.1$ & $1.10\pm 0.58$\\
$MS$ $0735.6+7421$ & $0.216$ & $722$ & $456\pm56$ & $1.90\pm 0.41$ & $1.60\pm 0.40$ & $1.78\pm 0.50$\\
$MS$ $0839.8+2938$ & $0.194$ & $664$ & $359\pm60$ & $1.13\pm 0.36$ & $ 0.74\pm 0.25$ & $0.59\pm 0.23$\\
$MS$ $0906.5+1110$ & $0.163$ & $578$ & $463\pm72$ & $1.25\pm 0.38$ & $1.9\pm1.1$ & $1.03\pm 0.24$\\
$MS$ $1006.0+1202$ & $0.221$ & $735$ & $392\pm169$ & $0.56\pm 0.93$ & $ 0.55\pm 0.72$ & $0.64\pm0.33$\\
$MS$ $1008.1-1224$ & $0.301$ & $921$ & $401\pm63$ & $0.49\pm 0.17$ & $1.19\pm 0.33$ & $0.43\pm 0.15$\\
$MS$ $1455.0+2232$ & $0.259$ & $828$ & $436\pm43$ & $1.90\pm 0.28$ & $1.50\pm 0.28$ & $1.29\pm 0.26$\\
$MS$ $2137.3-2353$ & $0.313$ & $946$ & $605\pm70$ & $3.30\pm 0.64$ & $4.2\pm1.1$ & $5.0\pm1.3$\\
$MS$ $J1157.3+5531$ & $0.081$ & $315$ & $495\pm175$ & $0.27\pm 0.21$ & $1.6\pm1.1$ & $0.36\pm 0.35$\\
$NGC$ $0507$ & $0.016$ & $69$ & $186\pm31$ & $0.069\pm0.026$ & $0.089\pm0.026$ & $0.0144\pm0.0062$\\
$NGC$ $4636$ & $ 0.003$ & $13$ & $171\pm19$ & $0.0246\pm0.0057$ & $0.071\pm0.015$ & $0.00356\pm 0.00092$\\
$NGC$ $5044$ & $ 0.009$ & $38$ & $159\pm30$ & $0.124\pm0.057$ & $0.055\pm0.018$ & $0.025\pm0.014$\\
\hline
  \end{tabular}
\end{table*}

\begin{table*}
\centering
\setcounter{table}{2}
\caption{continued.}
\renewcommand{\footnoterule}{}  
  \begin{tabular}{ | l c c c c c c |}
    \hline
    $cluster$ & $z$ & $D_{A}$ & $r_{2500}$ & $M_{gas}$ & $M_{tot}$ & $Y_{S}$\\
    $$ & $$ & $(Mpc)$ & $(kpc)$ & $(10^{13}M_{\odot})$ & $(10^{14}M_{\odot})$ & $(10^{-5}$ $Mpc^{2})$\\
    \hline
    \hline
$NGC$ $5846$ & $ 0.006$ & $24$ & $149\pm27$ & $0.0194\pm0.0070$ & $0.044\pm0.015$ & $0.00148\pm 0.00059$\\
$OPHIUCHUS$ & $0.028$ & $116$ & $294\pm39$ & $1.72\pm 0.58$ & $ 0.368\pm0.079$ & $2.8\pm1.1$\\
$PKS$ $0745-191$ & $0.103$ & $390$ & $579\pm52$ & $5.04\pm 0.79$ & $2.98\pm 0.56$ & $5.5\pm1.1$\\
$RBS$ $0533$ & $0.012$ & $52$ & $191\pm17$ & $0.075\pm0.015$ & $0.099\pm0.017$ & $0.0103\pm0.0015$\\
$RBS$ $0797$ & $0.354$ & $ 1026$ & $584\pm56$ & $4.23\pm 0.61$ & $4.04\pm 0.87$ & $6.9\pm1.4$\\
$RCS$ $J2327-0204$ & $0.200$ & $681$ & $582\pm91$ & $1.62\pm 0.49$ & $3.3\pm1.0$ & $2.41\pm 0.74$\\
$RXCJ$ $0331.1-2100$ & $0.188$ & $648$ & $561\pm75$ & $2.39\pm 0.65$ & $2.88\pm 0.73$ & $3.07\pm 0.76$\\
$RXC$ $J1023.8-2715$ & $0.096$ & $368$ & $622\pm152$ & $1.61\pm 0.65$ & $3.2\pm1.8$ & $2.0\pm1.2$\\
$RX$ $J0220.9-3829$ & $0.229$ & $755$ & $393\pm72$ & $0.49\pm 0.17$ & $1.02\pm 0.31$ & $0.32\pm 0.11$\\
$RX$ $J0232.2-4420$ & $0.284$ & $883$ & $478\pm92$ & $1.59\pm 0.55$ & $1.95\pm 0.72$ & $1.47\pm 0.64$\\
$RX$ $J0439.0+0715$ & $0.230$ & $758$ & $458\pm69$ & $1.32\pm 0.38$ & $1.68\pm 0.44$ & $1.09\pm 0.27$\\
$RX$ $J0439+0520$ & $0.208$ & $702$ & $512\pm79$ & $1.82\pm 0.54$ & $2.19\pm 0.76$ & $1.87\pm 0.65$\\
$RX$ $J0528.9-3927$ & $0.263$ & $837$ & $520\pm98$ & $2.49\pm 0.78$ & $2.5\pm1.1$ & $2.7\pm1.2$\\
$RX$ $J0647.7+7015$ & $0.584$ & $ 1362$ & $541\pm144$ & $3.6\pm1.5$ & $4.0\pm1.9$ & $5.8\pm2.5$\\
$RX$ $J0819.6+6336$ & $0.119$ & $443$ & $331\pm48$ & $0.535\pm 0.18$ & $ 0.55\pm 0.14$ & $0.266\pm0.081$\\
$RX$ $J1000.4+4409$ & $0.154$ & $551$ & $329\pm37$ & $0.365\pm0.089$ & $ 0.57\pm 0.13$ & $0.161\pm0.038$\\
$RX$ $J1022.1+3830$ & $0.049$ & $198$ & $243\pm32$ & $0.134\pm0.044$ & $ 0.209\pm0.051$ & $0.043\pm0.014$\\
$RX$ $J1130.0+3637$ & $0.060$ & $239$ & $232\pm35$ & $0.091\pm0.031$ & $ 0.178\pm0.050$ & $0.0199\pm0.0070$\\
$RX$ $J1320.2+3308$ & $0.037$ & $150$ & $205\pm29$ & $0.068\pm0.021$ & $ 0.122\pm0.032$ & $0.0114\pm0.0037$\\
$RX$ $J1347.5-1145$ & $0.451$ & $ 1189$ & $783\pm94$ & $10.3\pm2.1$ & $10.7\pm2.6$ & $34.7\pm8.8$\\
$RX$ $J1423.8+2404$ & $0.545$ & $ 1316$ & $578\pm65$ & $3.80\pm 0.67$ & $4.8\pm1.2$ & $7.0\pm1.8$\\
$RX$ $J1504.1-0248$ & $0.215$ & $720$ & $607\pm64$ & $4.60\pm 0.69$ & $3.83\pm 0.67$ & $5.5\pm1.0$\\
$RX$ $J1532.9+3021$ & $0.345$ & $ 1009$ & $499\pm107$ & $2.51\pm 0.73$ & $2.9\pm1.6$ & $2.39\pm 0.79$\\
$RX$ $J1539.5-8335$ & $0.073$ & $286$ & $612\pm160$ & $0.89\pm 0.43$ & $3.0\pm2.0$ & $1.31\pm 0.28$\\
$RX$ $J1720.1+2638$ & $0.164$ & $581$ & $496\pm42$ & $2.52\pm 0.38$ & $2.01\pm 0.32$ & $2.25\pm 0.41$\\
$RX$ $J1720.2+3536$ & $0.391$ & $ 1093$ & $455\pm43$ & $2.40\pm 0.41$ & $1.98\pm 0.34$ & $2.30\pm 0.45$\\
$RX$ $J1852.1+5711$ & $0.109$ & $412$ & $327\pm38$ & $0.363\pm0.089$ & $ 0.57\pm 0.15$ & $0.152\pm0.038$\\
$RX$ $J2129.6+0005$ & $0.235$ & $770$ & $696\pm301$ & $3.2\pm1.6$ & $6.0\pm4.9$ & $6.4\pm4.0$\\
$SC$ $1327-312$ & $0.053$ & $213$ & $307\pm34$ & $0.343\pm0.098$ & $ 0.425\pm0.082$ & $0.156\pm0.037$\\
$SERSIC$ $159-03$ & $0.058$ & $232$ & $407\pm116$ & $0.58\pm 0.28$ & $ 0.85\pm 0.49$ & $0.284\pm0.079$\\
$SS2B153$ & $0.0190000$ & $78$ & $315\pm98$ & $0.108\pm0.049$ & $ 0.71\pm 0.40$ & $0.071\pm0.051$\\
$UGC$ $03957$ & $0.034$ & $140$ & $281\pm20$ & $0.278\pm0.043$ & $ 0.323\pm0.051$ & $0.083\pm0.014$\\
$UGC$ $12491$ & $0.017$ & $73$ & $386\pm81$ & $0.31\pm 0.14$ & $ 0.78\pm 0.29$ & $0.23\pm 0.12$\\
$ZwCl$ $0857.9+2107$ & $0.235$ & $770$ & $367\pm72$ & $0.57\pm 0.20$ & $ 0.83\pm0.30$ & $0.282\pm0.086$\\
$ZWCL$ $1215$ & $0.075$ & $294$ & $448\pm69$ & $1.27\pm 0.47$ & $1.31\pm 0.36$ & $1.23\pm 0.45$\\
$ZWCL$ $1358+6245$ & $0.328$ & $976$ & $479\pm151$ & $1.9\pm1.1$ & $1.88\pm1.31$ & $2.9\pm2.6$\\
$ZWCL$ $1742$ & $0.076$ & $296$ & $539\pm195$ & $0.70\pm 0.35$ & $1.9\pm1.5$ & $0.54\pm 0.24$\\
$ZWCL$ $1953$ & $0.380$ & $ 1074$ & $531\pm82$ & $2.95\pm 0.84$ & $3.0\pm 0.86$ & $3.46\pm 0.95$\\
$ZWCL$ $3146$ & $0.290$ & $897$ & $486\pm48$ & $3.65\pm 0.66$ & $2.17\pm 0.41$ & $3.58\pm 0.73$\\
$ZWICKY$ $2701$ & $0.210$ & $707$ & $426\pm59$ & $1.35\pm 0.31$ & $1.29\pm 0.34$ & $0.85\pm 0.22$\\
\hline
  \end{tabular}
  \label{Tab_all} 
\end{table*}
\label{lastpage}

\end{document}